\documentclass[12pt,aps,tightenlines,notitlepage,superscriptaddress]{revtex4-1}

\usepackage{amsmath}
\usepackage{amssymb}
\usepackage{bm}
\usepackage[mathscr]{eucal}
\usepackage{theorem}
\usepackage{graphicx}
\usepackage{psfrag}
\usepackage{subfigure}
\usepackage{color}
\usepackage{wasysym}
\include{graphix}
\usepackage{framed}
\usepackage{enumitem}%
\usepackage{lipsum}
\usepackage{float}
\usepackage{MnSymbol}
\usepackage{empheq}
\usepackage{fancybox}

\usepackage{float}
\usepackage{amscd}
\usepackage{wasysym}

\graphicspath{{figs/}}

\usepackage{tikz}
\newcommand{\Depth}{8}
\newcommand{\Height}{4}
\newcommand{\Width}{2}

\newcommand{\vecA}{\textbf{\textit{A}}}
\newcommand{\vecx}{\textbf{\textit{x}}}
\newcommand{\vecu}{\textbf{\textit{u}}}
\newcommand{\veck}{\textbf{\textit{k}}}
\newcommand{\silly}{a}
\newcommand{\mathd}{\mathrm{d}}
\newcommand{\mathe}{\mathrm{e}}
\newcommand{\imag}{\mathrm{i}}

\newcommand{\simples}{S-TPLS$\text{ }$}
\newcommand{\simplesend}{S-TPLS}

\begin{document}

\bibliographystyle{apsrev}

\title{Simplified TPLS as a learning tool for high-performance computational fluid dynamics}

\author{James Fannon\footnote{Present address: Department of Mathematics and Statistics, University of Limerick, Ireland}}
\affiliation{School of School of Mathematics and Statistics, University College Dublin, Belfield, Dublin 4}
\affiliation{School of Physics, University College Dublin, Belfield, Dublin 4, Ireland}

\author{Jean-Christophe Loiseau}
\affiliation{Department of Mechanics, KTH Royal Institute of Technology, SE-100 44 Stockholm, Sweden}

\author{Prashant Valluri}
\affiliation{Institute for Materials and Processes, School of Engineering, The University of Edinburgh, Edinburgh EH9 3BF, UK}

\author{Iain Bethune}
\affiliation{EPCC, The University of Edinburgh, Edinburgh EH9 3FB, UK}

\author{Lennon \'O N\'araigh}
\email{onaraigh@maths.ucd.ie}
\affiliation{School of Mathematics and Statistics, University College Dublin, Belfield, Dublin 4}
\affiliation{Complex \& Adaptive Systems Laboratory, University College Dublin, Belfield, Dublin 4, Ireland}

\date{\today}

\begin{abstract}
We introduce a modified and simplified version of the pre-existing fully parallelized three-dimensional Navier--Stokes flow solver known as TPLS.  We demonstrate how the simplified version can be used as a pedagogical tool for the study of computational fluid dynamics and parallel computing.
TPLS is at its heart a two-phase flow solver, and uses calls to a range of external libraries to accelerate its performance.   However, in the present context we narrow the focus of the study to basic hydrodynamics and parallel computing techniques, and the code is therefore simplified and modified to simulate pressure-driven single-phase flow in a channel, using only relatively simple Fortran 90 code with MPI parallelization, but no calls to any other external libraries.
The modified code is analysed in order to both validate its accuracy and investigate its scalability up to 1000 CPU cores.
Simulations are performed for several benchmark cases in pressure-driven channel flow, including a turbulent simulation, wherein the turbulence is incorporated via the large-eddy simulation technique.
The work may be of use to advanced undergraduate and graduate students as an introductory study in computational fluid dynamics, while also providing insight for those interested in more general aspects of high-performance computing.
\end{abstract}

\maketitle

\section{Introduction}
\label{sec:intro}

TPLS (Two Phase Level Set) is an accurate, highly efficient two-phase Navier--Stokes (NS) solver, parallelized and scalable up to 1000s of CPU cores~\cite{scott2013}.   The code is available for research-level production runs as open-source software~\cite{tpls_sourceforge}.
The development of TPLS was motivated by open questions in the two-phase flow literature (e.g. Reference~\cite{lennon_1}), which may not be of concern to a student seeking to develop basic skills in Computational Fluid Dynamics (CFD).  Therefore, the aim of the present work is to introduce a simplified (single-phase) version of TPLS to expose students to high-performance computing through the medium of some fairly classical problems in hydrodynamics -- namely the supercritical instability of single-phase pressure-driven channel flow, and fully-developed turbulence in the same.   In this way, the present work illustrates how contemporary research can inform the understanding of physics at university level.
In this introduction, we outline some particular features of TPLS, as well as placing the above physical problems in the context of the broader literature on computational fluid dynamics.  We also take care to present the computational methodology in the broader context of CFD education.

The full research-level version of TPLS~\cite{tpls_sourceforge} is unique in several aspects.  The TPLS solver has been custom-built for supercomputing architectures with large scale simulations of complicated interfacial fluid flows in mind.  The full research-level version exploits parallel libraries that are typically available at supercomputing centres, such as PETSc~\cite{petsc} and NetCDF~\cite{netcdf} -- see Reference~\cite{bethune2015}.  In order to present a highly portable version of the code to students,
the methodology presented in this work strips back some of this complexity.  The result is a (single-phase) code capable of being run on desktop computers, clusters, and supercomputers -- referred to throughout this work as S-TPLS (for simplified-TPLS).
Some degradation in the code's performance is expected as a result of this simplification, but the payoff in terms of simplicity and portability is considerable.  Performance analysis of the simplified code is addressed in this paper in Section~\ref{sec:performance}.

Concerning the physical problems discussed herein, simulation and modelling of turbulent flow provides an attractive problem for students, with the beauty, complexity and infamy of turbulence itself acting as a strong combination of motivating factors.
Parametrization of the unresolved eddies in the simulation is a highly non-trivial problem, and a brief literature review concerning LES and turbulence simulation more generally is appropriate here, before the methodology and the results of the present study are presented.
 Heuristically, a turbulent flow is characterized by the non-linear development of eddies (particular patches of fluid illustrating coherent motion) on a wide range of length scales, and velocity vector and pressure fields which are subject to random fluctuations in both time and space, although more formal definitions can be given in terms of some underlying properties \cite{berselli,Abbott}.
The impasse in terms of an analytical understanding of turbulence (i.e. the open problem of existence and smoothness of solutions to the NS equations~\cite{CMI}) has motivated a numerical approach to the problem in order to gain further insight. However, simulating turbulent flow presents two main problems of its own:
\begin{enumerate}
\item Computational restrictions -- An accurate simulation must resolve the fluid motion across all pertinent length scales, which leads to the requirement that the total number of grid points $N_{T}$ needed in the simulations scales as~\cite{Pope}
\begin{equation*}
N_{T} \sim Re_{\ell}^{9/4}, \qquad Re_{\ell}=\frac{\rho U \ell}{\mu},
\end{equation*}
where $Re_{l}$ is the Reynolds number based on the large-scale eddies, $\rho$ is the fluid density, $U$ is a characteristic velocity scale, and $\mu$ is the dynamic viscosity of the fluid. Many physical problems of interest have large values of $Re_{\ell}$, the simulation of which has a huge computational cost.

\item The closure problem --  One may try to circumvent the first problem and apply an averaging process to the NS equations and solve for the averaged properties of the flow, which may be of interest, instead of the full velocity and pressure fields. However, these averaged equations -- known as the Reynolds Averaged Navier-Stokes (RANS) equations -- contain additional stress terms which have no known form \textit{a priori} and thus must be modelled.
\end{enumerate}

The LES technique (see Section~\ref{sec:problem}) represents a half-way mark between a direct numerical simulation (DNS) and a RANS simulation. However, a large-scale LES still presents a considerable computational challenge and thus such a simulation is generally split between many CPU cores. In the case of this work, both a local cluster and a supercomputer (Fionn, ICHEC) were used for the simulations, with the local cluster proving  adequate for almost all the requirements of the work.
In this way, the LES problem provides a natural introduction to  the area of parallel computing, an area of increasing importance for graduate and research work in mathematics and physics.  Indeed, using a simplified version of TPLS that can be ported across a variety of architectures, it is possible to introduce in a non-trivial way the topic of high-performance computing to undergraduates in mathematics and the physical sciences in a single semester.
The discussion naturally leads to more advanced topics, such as
performance analysis and parallel efficiency of the computational code.

In the present introduction, it is also a useful exercise to show how our computational methodology fits inside the broader context of CFD education.   While there are several commercial CFD solvers taught in current undergraduate and graduate science curricula the world over, the emphasis is more on a `black-box' approach with focus on handling the software for standard problems including turbulent flow in a channel/pipe. Typically, these standard CFD courses are delivered predominantly as electives in the graduating years, with students having taken fundamental courses in fluid mechanics courses in the earlier years. However, more often, there is a mismatch in the learning objectives and delivery aspects for both these courses - with a CFD course appearing to adopt a more black-box type of approach - leading to less emphasis on the fundamental physics governing the flows. Moreover, the common commercial software being used for these courses provide only a graphical user interface (GUI), with little or no access to the solvers being used. While this is rightly so to protect IP and also ensure stability of the code as it is less amenable to tinkering, the students only have experience in treating a flow problem using a GUI that is highly specific to the software available in their curricula. This is particularly true for complex flow phenomena such as turbulence where a RANS-type models with specified eddy viscosity models are predominantly promoted as the method of choice in commercial flow solvers - with little or no means of introducing even rudimentary levels of customisation. Also, a typical turbulent flow example within vendor-prescribed tutorials (in commercial codes) would be based on a `steady-state two-dimensional or axisymmetric' solution - giving a completely incorrect impression to the students that turbulence is steady and two-dimensional. While it is incumbent on the instructor to correct this impression - examples of such fictitious flows are too ubiquitous (in almost every commercial solver and even old peer-reviewed articles) for students to ignore them. Of course, this is gradually changing with availability of LES methods in some solvers which will only work with transient approaches - but again only with rigidly specified models for Reynolds stresses to account for sub-grid turbulence. A further problem is the terminology used in commercial solvers, where they prescribe a DNS approach as a `laminar' model.  Undergraduate students are then exposed to a risky solecism -- believing that a DNS approach is `unsuitable' for turbulent flows.  While many university departments have facilities for local computing clusters, the standard parallelisation schemes in commercial CFD solvers renders their efficiency sub-optimal. This means that the undergraduate students are only exposed to approximate `simple' single-core problems for `laminar' (low $Re$) or `steady-state 2D/axisymmetric RANS-type turbulent' (high $Re$) flows and never get hands-on experience with high-performance parallelised computing for transient real-life flows (which is mostly exclusive to senior graduate students or post-doctoral researchers).

With highly efficient parallelisation schemes and a transient DNS based solver with customisable LES models, S-TPLS offers a step-change in delivery of CFD instruction using state-of-art tools. S-TPLS through its simple and openly-available Fortran routines, also provides a unique opportunity for rigorous customisation based on their project or assignment problem at hand. This is a complete departure from a black-box only to a full-tinkering approach, giving the students freedom (for example) to introduce new LES models, or customise boundary conditions.
The LES model currently introduced in this paper alongside S-TPLS solver makes it also computationally feasible across a wide range of platforms.

This article is organized as follows.
The mathematical formulation of the problem is presented along with the LES techniques in Section~\ref{sec:problem}.
The \simples solver is introduced as a means of solving the problem in Section~\ref{sec:tpls}.
A stringent validation method for the solver is presented in Section~\ref{sec:validation}, while the performance of the solver is rigorously analysed in Section~\ref{sec:performance}.  Results of the LES are described in Section~\ref{sec:results}.  Finally, concluding remarks are provided in Section~\ref{sec:conc}, together with a brief discussion concerning
the application of this work as a learning tool for CFD.

\section{Problem statement}
\label{sec:problem}

\subsection{Navier--Stokes equations}

We consider an incompressible, Newtonian fluid confined in a channel geometry $\Omega = \left[0,L_x\right] \times \left[0,L_y\right] \times \left[0,L_z\right]$ subject to a constant  negative pressure gradient $dp/dx$ in the $x$ direction. In the absence of gravity and  other external forces, the non-dimensional NS equations can be written in component form as
%
\begin{subequations}
\begin{eqnarray}
\frac{\partial u_{i}}{\partial t} + u_{j} \frac{\partial u_{i}}{\partial x_{j}} = -\frac{\partial p}{\partial x_{i}}+\frac{1}{Re_*}\frac{\partial^2 u_{i}}{\partial x_{j} \partial x_{j}}, \label{eq:NSE_momentum_nond} \\
\frac{\partial u_i}{\partial x_i} = 0 \label{eq:NSE_continuity_nond},
\end{eqnarray}%
\label{eq:NS_all}%
\end{subequations}%
where the Einstein summation convention is used, $u_i$ denotes the
velocity-field components $\textbf{\textit{u}} = (u,v,w) \equiv (u_1,u_2,u_3)$, and $p$ denotes the pressure field.  The quantity $Re_*$ is the Reynolds number $Re_*=\rho U L_z/\mu$ based on the friction velocity
\begin{equation}
U=\sqrt{\frac{L_z}{2\rho}\left|\frac{dp}{dx}\right|}.
\end{equation}
In this non-dimensional scheme, the non-dimensional channel height is set to unity, $L_z\rightarrow 1$.

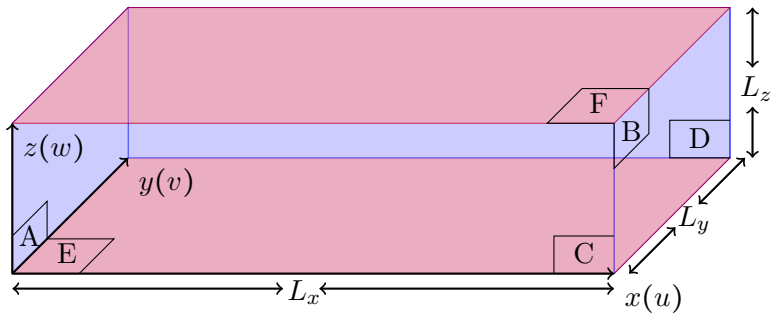
\begin{figure}[htb]
\begin{center}
\begin{tikzpicture}
\coordinate (O) at (0,0,0);
\coordinate (A) at (0,\Width,0);
\coordinate (B) at (0,\Width,\Height);
\coordinate (C) at (0,0,\Height);
\coordinate (D) at (\Depth,0,0);
\coordinate (E) at (\Depth,\Width,0);
\coordinate (F) at (\Depth,\Width,\Height);
\coordinate (G) at (\Depth,0,\Height);

\draw[red,fill=red!40] (O) -- (C) -- (G) -- (D) -- cycle;
\draw[blue,fill=blue!20] (O) -- (A) -- (E) -- (D) -- cycle;
\draw[blue,fill=blue!20] (O) -- (A) -- (B) -- (C) -- cycle;
\draw[blue,fill=blue!20,opacity=0.4] (D) -- (E) -- (F) -- (G) -- cycle;
\draw[blue,fill=blue!20,opacity=0.4] (C) -- (B) -- (F) -- (G) -- cycle;
\draw[red,fill=red!40,opacity=0.6] (A) -- (B) -- (F) -- (E) -- cycle;

\draw[thick,->] (0,0,\Height) -- (0,\Width,\Height) node[anchor=north west]{$z(w)$};
\draw[thick,->] (0,0,\Height) -- (\Depth,0,\Height) node[anchor=north west]{$x(u)$};
\draw[thick,->] (0,0,\Height) -- (0,0,0) node[anchor=north west]{$y(v)$};

\draw  (0,0,\Height) -- (0,\Width/4,\Height) -- (0,\Width/4,7*\Height/10) -- (0,0,7*\Height/10);%
\node [above] at (0,0,8.5*\Height/10) {A};

\draw  (0,0,\Height) -- (\Depth/9,0,\Height) -- (\Depth/9,0,7*\Height/10) -- (0,0,7*\Height/10);
\node [above] at (\Depth/11,0,\Height) {E};

\draw  (\Depth,\Width,\Height) -- (8*\Depth/9,\Width,\Height) -- (8*\Depth/9,\Width,7*\Height/10) -- (\Depth,\Width,7*\Height/10);
\node [above] at (9.75*\Depth/10,\Width,\Height) {F};

\draw  (\Depth,\Width,\Height) -- (\Depth,7*\Width/10,\Height) -- (\Depth,7*\Width/10,7*\Height/10) -- (\Depth,\Width,7*\Height/10);
\node [above] at (\Depth,7*\Width/10,8.5*\Height/10) {B};

\draw  (\Depth,0,\Height) -- (9*\Depth/10,0,\Height) -- (9*\Depth/10,\Width/4,\Height) -- (\Depth,\Width/4,\Height);
\node [above] at (9.5*\Depth/10,0,\Height) {C};

\draw  (\Depth,0,0) -- (9*\Depth/10,0,0) -- (9*\Depth/10,\Width/4,0) -- (\Depth,\Width/4,0);
\node [above] at (9.5*\Depth/10,0,0) {D};

\draw[thick,<->] (\Depth,-0.2,\Height) -- (2.55*\Depth/5,-0.2,\Height);
\draw[thick,<->] (2.25*\Depth/5,-0.2,\Height) -- (0,-0.2,\Height);

\node [left] at (2.65*\Depth/5,-0.25,\Height) {$L_x$};

\draw[thick,<->] (\Depth+0.2,0,\Height) -- (\Depth+0.2,0,3.0*\Height/5);
\draw[thick,<->] (\Depth+0.2,0,2.0*\Height/5) -- (\Depth+0.2,0,0);
\node [right] at (\Depth,0,2.7*\Height/5) {$L_y$};

\draw[thick,<->] (\Depth+0.3,0,0) -- (\Depth+0.3,1.7*\Width/5,0);
\draw[thick,<->] (\Depth+0.3,3*\Width/5,0) -- (\Depth+0.3,\Width,0);
\node [right] at (\Depth,2.3*\Width/5,0) {$L_z$};

\end{tikzpicture}
\caption{\emph{Computational domain $\Omega = \left[0,L_x\right] \times \left[0,L_y\right] \times \left[0,L_z=1\right]$. Periodic boundary conditions are used in the streamwise (faces AB) and spanwise (faces CD) directions, while no-slip boundary conditions are used on faces $E$ and $F$.}}
\label{fig: Comp domain}
\end{center}
\end{figure}

The computational domain $\Omega$ is illustrated in Figure~\ref{fig: Comp domain}. The velocity field is subject to periodic boundary conditions in the streamwise ($x$) and spanwise ($y$) directions, while no-slip boundary conditions are enforced at the walls i.e. $\textbf{\textit{u}}(x,y,z=0,t)=0=\textbf{\textit{u}}(x,y,z=1,t)$.  Periodic boundary conditions in the streamwise and spanwise directions are also applied to the pressure field, modulo the constant pressure drop $dp/dx$ applied in the $x$-direction.  This amounts to a constant linear forcing of the flow, which sustains the mean flow in the $x$-direction.
The use of periodic boundary conditions in the streamwise and spanwise directions is appropriate provided that the size of the computational domain in these directions is such that the largest-scale eddies in the flow can be accommodated~\cite{Kim}.  This is checked \textit{a posteriori} when simulation results are postprocessed.
%
%

\subsection{LES technique}

The LES method of simulating turbulent flow has been in existence for over 40 years \cite{Deardoff} and is covered in detail in many textbooks \cite{Abbott,Pope,Davidson}. Essentially, the idea underpinning the LES method is the following; instead of trying to resolve the entire velocity and pressure fields in a simulation, one solves for the so-called filtered velocity and pressure fields which describe the fluid motion exactly down to a given filter width. Small-scale structures which exist below this filter width are thus not resolved in the LES, but their effect on the rest of the flow must be modelled. This process allows for the large-scale structures present in the flow to be resolved while avoiding the high computational cost associated with resolution of small-scale structures.

In order to obtain the dynamic equations which describe the filtered velocity and pressure fields, we apply a filtering process to Equation~\eqref{eq:NS_all} by forming a convolution with a  filter function $\mathcal{G}\left(\textbf{\textit{y}}\right)$. The filtered equations of motion one obtains  are given by
\begin{subequations}
\begin{eqnarray}
\frac{\partial \overline{u_i}}{\partial t} + \overline{u_j}\frac{\partial \overline{u_i}}{\partial x_j} = -\frac{\partial \overline{p}}{\partial x_i} + \frac{1}{Re_*}\frac{\partial^2 \overline{u_i}}{\partial x_j \partial x_j} - \frac{\partial \overline{\tau_{ij}}}{\partial x_j} \label{eq: Filterned Momentum Eq} \\
\frac{\partial \overline{u_i}}{\partial x_i} = 0, \label{eq: Filtered Continutiy eq}
\end{eqnarray}%
\label{eq:filtered}%
\end{subequations}%
where the over-bar refers to the filtered fields. We define the residual stress tensor
\begin{equation}
\overline{\tau_{ij}} = \overline{u_i u_j}-\overline{u_i} \hspace{2pt} \overline{u_j},
\label{eq: SGS term}
\end{equation}
which is unknown as it contains the unknown velocity components $u_i$. In order to close this set of equations, we employ the standard Smagorinsky model for $\overline{\tau_{ij}}$ using
\begin{equation}
\overline{\tau_{ij}}=-2 \nu_{t} \overline{\textbf{\textit{s}}_{ij}}, \qquad \overline{\textbf{\textit{s}}_{ij}}= \frac{1}{2}\left(\frac{\partial \overline{u_i}}{\partial x_j} + \frac{\partial \overline{u_j}}{\partial x_i}\right),
\label{eq: Smag model}
\end{equation}
where $\nu_{t}$ is the so-called eddy viscosity, given by
\begin{equation}
\nu_{t}\left(\textbf{\textit{x}},t\right)= C_{S}^2 \Delta^2 \left|\overline{\textbf{\textit{s}}}\right|, \qquad \left|\overline{\textbf{\textit{s}}} \right| = \sqrt{2 \left(\overline{\textbf{\textit{s}}_{ij}}\right) \left(\overline{\textbf{\textit{s}}_{ij}}\right)},
\end{equation}
where $C_{S}$ is the dimensionless Smagorinsky coefficient and $\Delta$ is the length scale of the largest unresolved eddy present in the turbulent flow. In this article, we use $C_{S}=0.1$~\cite{Davidson} and $\Delta = 2\left(\Delta x \Delta y \Delta z \right)^{1/3}$~\cite{Abbott}, where $(\Delta x,\Delta y,\Delta z)$ denote the grid spacings in the $x,y$ and $z$ directions, respectively. Finally, we also incorporate a near-wall modelling term of the form
\begin{equation}
\phi_{w}\left(z\right)=\begin{cases}
 \left\{ 1-\exp\left[\left(\frac{-z Re_*}{A}\right)\right]^p \right\} ^q& z\leq \frac{1}{2}\\
\left\{ 1-\exp\left[\frac{-\left(1-z\right)Re_*}{A}\right]^p\right\}^q& z \geq \frac{1}{2}
\end{cases}
\end{equation}
as an additional pre-factor for the length scale $\Delta$ in order to take into account the increased turbulence production close to the walls at $z=0,1$. In this article, we take the values $p=q=1$ and $A=25$, in keeping with the standard Van Driest components~\cite{VanDriest}. Thus, the eddy viscosity term becomes $\nu_{t} = \left(C_{S}\Delta \phi_w\right)^2 \left|\overline{\textbf{\textit{s}}}\right|$. Incorporating the Smagorinsky model as given by Equation~\eqref{eq: Smag model} into the filtered momentum equation~\eqref{eq: Filterned Momentum Eq} yields
\begin{subequations}
\begin{eqnarray}
\frac{\partial \overline{u_i}}{\partial t} + \overline{u_j}\frac{\partial \overline{u_i}}{\partial x_j} = -\frac{\partial \overline{p}}{\partial x_i} + \frac{\partial}{\partial x_j}\left[\left(\frac{1}{Re_*} + \nu_{t}\right)\left(\frac{\partial \overline{u_i}}{\partial x_j} + \frac{\partial \overline{u_j}}{\partial x_i}\right) \right], \label{eq:NS_momentum_LES} \\
\frac{\partial \overline{u_i}}{\partial x_i} = 0,\label{eq:NS_continuity_LES}
\end{eqnarray}%
\label{eq:NS_LES}%
\end{subequations}%
which are immediately reminiscent of the dimensionless NS equations~\eqref{eq:NS_all}. The key difference is that incorporating the Smagorinsky model effectively introduces a non-constant viscosity term $\nu_{T} =  Re_*^{-1} + \nu_t$ into the equations.

\section{\simples solver}
\label{sec:tpls}

With the model problem established, attention now turns to solving Equation~\eqref{eq:NS_LES} numerically, which is done by the \simplesend code. As stated previously, the full research-level version of TPLS solves the NS equations for two interacting fluids. Since we are only interested in single-phase  flow in this article, the code  used herein is simplified by stripping away the second fluid to create a single-phase NS solver -- hence, \simplesend.  However, since a two-phase flow has a non-constant viscosity in the levelset formalism~\cite{kang2000boundary}, the generic structure of the algorithms in TPLS is ideally suited to handling a non-constant viscosity such as that in Equation~\eqref{eq:NS_LES}.   In presenting the computational methodology below, we omit the overbar over the spatially-filtered velocity and pressure fields, since the context of the presentation indicates when we are dealing with an LES simulation.

\subsection{Discretization}

\begin{figure}
\centering
\begin{tikzpicture}
\draw [step=2cm,gray,very thin,dotted] (0.1,0.1) grid (5.9,5.9);
\draw [thick,red,->] (3.75,3) -- (4.25,3) node[anchor=north,red]{$u(v)$};
\draw [thick,red,->] (1.75,3) -- (2.25,3) node[anchor=north,red]{$u(v)$};
\draw [thick,blue,->] (3,1.75) -- (3,2.25) node[anchor=west,blue]{$w$};
\draw [thick,blue,->] (3,3.75) -- (3,4.25) node[anchor=west,blue]{$w$};
\foreach \Point in {(3,3),(1,3),(5,3),(3,5),(3,1)}{
	\node at \Point {\textbullet};
}
\node [above] at (3,3) {$p_{\ell(m),n}$};
\node [above] at (3,1) {$p_{\ell(m),n-1}$};
\node [above] at (3,5) {$p_{\ell(m),n+1}$};
\node [above] at (1,3) {$p_{\ell-1(m-1),n}$};
\node [above] at (5,3) {$p_{\ell+1(m+1),n}$};

\draw [thick,->] (0,0) -- (5.9,0) node[anchor=north west]{$x(y)$};
\draw [thick,->] (0,0) -- (0,5.9) node[anchor=south east]{$z$};
\end{tikzpicture}
\caption{\emph{Schematic of a two dimensional MAC grid, in the $xz$ or equivalently $yz$ plane, indicating the location of pressure and velocity values.}}
\label{fig: General MAC 2D schematic}
\end{figure}
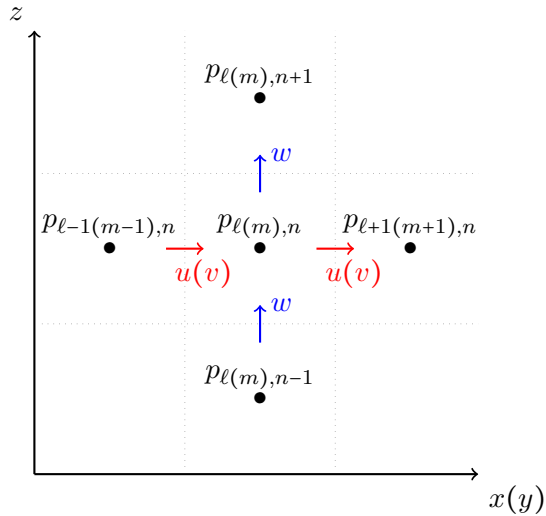

The  domain $\Omega$ is discretized in a uniform manner using a finite-volume scheme.  Thus, the computational mesh has  $(N_x,N_y,N_z)$ grid points in the $(x,y,z)$ directions, with uniform grid spacings  $(\Delta x,\Delta y, \Delta z)$.   A three-dimensional marker-and-cell (MAC) grid~\cite{Harlow} is used for this purpose. The idea is illustrated in two dimensions in Figure~\ref{fig: General MAC 2D schematic}, but one can immediately envision its extension to three dimensions.  The variables $u$, $v$, $w$, and $(p,\nu_T)$  have their own computational grids. The values for $p$ and $\nu_T$ are stored at the cell centres while velocity components are stored at the cell faces. As such, the use of a MAC grid allows spatial derivatives to be approximated numerically as a difference taken between two cell faces, accurately taking into account the momentum flux between cells.  Additionally, the use of the MAC grid for the incompressible fluid equations~\eqref{eq:NS_all} stabilizes the code numerically against the checkerboard instability~\cite{Harlow}.

Concerning the temporal discretization of the code, the momentum equation~\eqref{eq:NS_momentum_LES} is solved using the so-called projection method~\cite{Chorin}. In this method, the pressure term is first omitted from the momentum equation which is then solved for at an intermediate half-step $\textbf{\textit{u}}^*$ i.e.
\begin{equation}
\frac{\textbf{\textit{u}}^* - \textbf{\textit{u}}^n}{\Delta t} + \textbf{\textit{u}}\cdot \nabla \textbf{\textit{u}} = \nabla \cdot \left(\nu_T \left(\nabla\textbf{\textit{u}} + \nabla \textbf{\textit{u}}^{T}\right)\right) \label{eq:momentum half step}
\end{equation}
where $\textbf{\textit{u}}^{n}$ is the velocity field at the $n$-th time step, $\Delta t$ is the time step, and the total viscosity $\nu_T$ is  a time- and space-dependent variable.  The viscous derivative is split into terms which appear more convective and others which appear more diffusive in nature. For example, by expanding the term on the right hand side of Equation~\eqref{eq:momentum half step}, one obtains terms such as (again for $j=1,2,3$)
\begin{equation}
\frac{\partial}{\partial x_i}\left(\nu_T \frac{\partial u_i}{\partial x_j}\right),
\label{eq:more_conv}
\end{equation}
which we recognise to be convective in nature, and terms which appear more diffusive
\begin{equation}
\frac{\partial}{\partial x_i}\left(\nu_T\frac{\partial u_j}{\partial x_i}\right).
\label{eq:more_diff}
\end{equation}
%
The first of these terms (i.e. Equation~\eqref{eq:more_conv})  is discretized in time using a third-order Adams--Bashforth scheme~\cite{Boyd}.  The same temporal discretization scheme is used for the convective derivative $\vecu\cdot\nabla\vecu$.  A Crank--Nicolson scheme~\cite{garcia2000} is used for the more diffusion-like terms in Equation~\eqref{eq:more_diff}. Equation~\eqref{eq:momentum half step} is then solved numerically for $\textbf{\textit{u}}^*$, which is then used to obtain the velocity field at the next time step $\textbf{\textit{u}}^{n+1}$ by reintroducing the pressure term and making the following correction to $\textbf{\textit{u}}^*$
\begin{equation}
\frac{\textbf{\textit{u}}^{n+1}-\textbf{\textit{u}}^*}{\Delta t} = -\nabla p^{n+1}.
\label{eq:project u_p_1}
\end{equation}
Taking the divergence of both sides and using the incompressibility requirement on the field $\vecu^{n+1}$ i.e. $\nabla \cdot \vecu^{n+1}=0$ one obtains
\begin{equation}
\nabla ^2 p^{n+1} = \left(\frac{\nabla \cdot \vecu^*}{\Delta t}\right),
\label{eq:presdiv}
\end{equation}
i.e. Poisson's equation. Once $p^{n+1}$ is determined, this allows us to solve for $\textbf{\textit{u}}^{n+1}$ via Equation~\eqref{eq:project u_p_1}.

\subsection{Error analysis, code stability, and convergence}

Centred differences are used throughout the code in evaluating spatial derivatives.  The centred differences are implemented on the MAC grid.  Thus, the error associated with approximating $\partial \phi/\partial x$ by a finite difference is $O(\Delta x^2)$, where $\phi$ is a generic field variable.   Concerning the convective derivative $\vecu\cdot\nabla\vecu$, a more accurate treatment  would involve upwinding and possibly accounting for local sharp changes in the velocity field (for example, using a (weighted) ENO scheme~\cite{shu1988efficient,liu1994weighted}, as in Reference~\cite{kang2000boundary}).  However, the centred differences are shown in the simulation results to be adequate for our purposes.

The temporal discretization involves both a Crank--Nicholson discretization for the diffusion and a third-order Adams--Bashforth step for the convective derivative.  The schemes introduce an error over the course of the entire simulation that is $O(\Delta t^n)$, with $n=2$ for the Crank--Nicholson part~\cite{Boyd} and $n=4$ for the Adams--Bashforth part~\cite{durran1991third}.
The explicit  treatment of the convective derivative in the temporal discretization introduces a CFL stability constraint on the code~\cite{Boyd}.  A conservative rule of thumb to maintain stability in the simulations is found to be
\begin{equation}
U_{\mathrm{max}}\Delta t\leq 0.1\Delta x,
\label{eq:cfl_cond}
\end{equation}
where $U_\mathrm{max}$ is the maximum streamwise velocity.
Finally, the simulations are checked for convergence of the solutions with respect to the grid spacings $\Delta x,\Delta y, \Delta z$, and the stepsize $\Delta t$.  The turbulence simulation requires additional care in selecting converged values of the grid spacings, as the grid spacings must be small so that the LES captures enough of the energy-containing motions so as to produce accurate statistics for the mean flow.

\subsection{Domain decomposition and output data}

\begin{figure}
	\centering
		\includegraphics[width=0.6\textwidth]{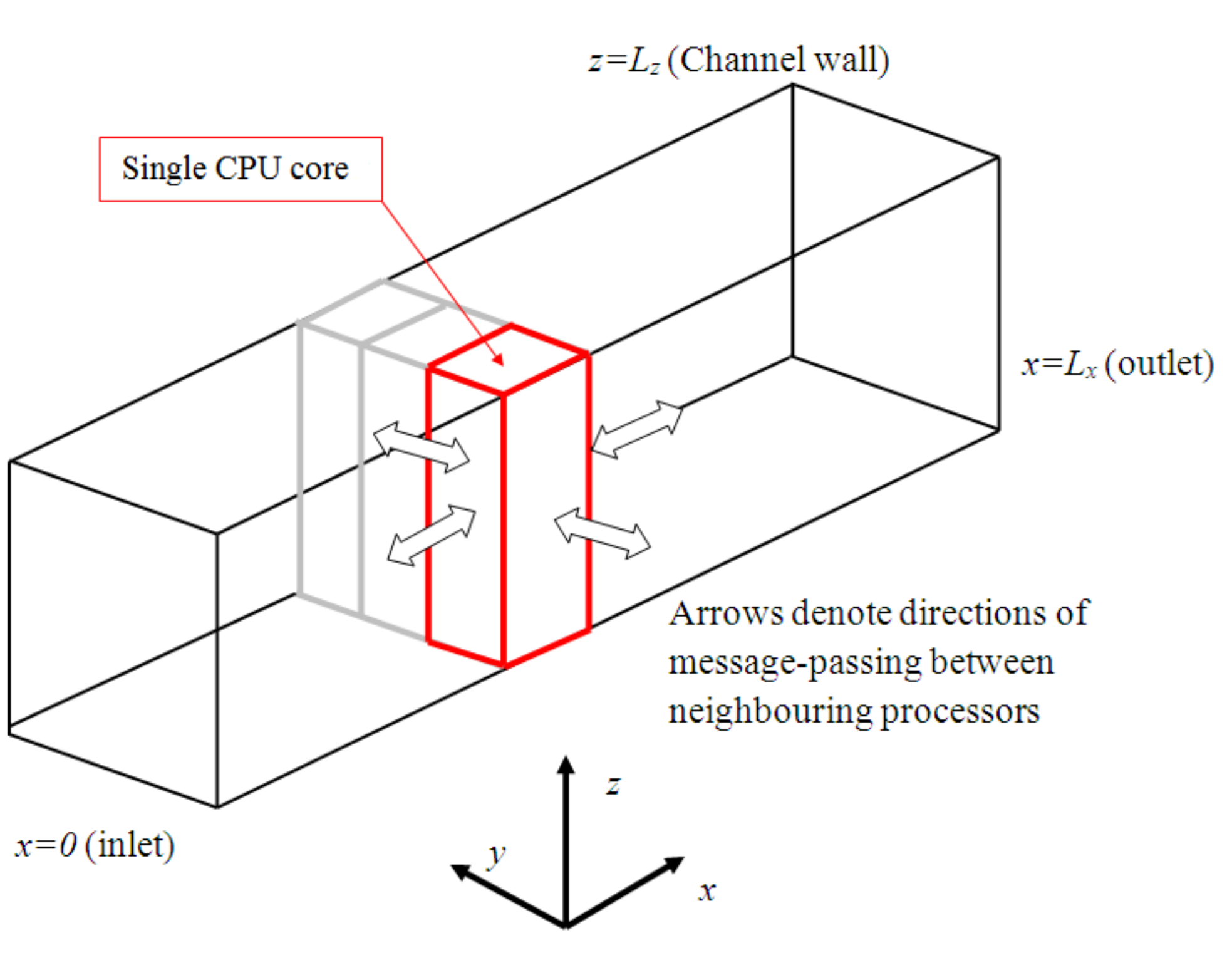}
		\caption{Schematic diagram showing the domain decomposition for the parallel code.  Geometrically, each sub-domain is an elongated box spanning the entire wall-normal direction.  The Navier--Stokes equations are solved on each sub-domain on a particular CPU core.  Information is passed between the sub-domains using the MPI library.}
	\label{fig:schematic2}
\end{figure}

The presented code is implemented in Fortran 90 with MPI parallelization using a two-dimensional domain decomposition in the $x$- and $y$-directions.  This means that the computational domain is split into several smaller sub-domains (as in Figure~\ref{fig:schematic2}),
 and the numerical computations corresponding to each sub-domain are performed on separate CPU cores.  Each sub-domain spans the entire wall-normal direction, such that the decomposition is two-dimensional.  The reason for this is that the wall-normal boundary conditions are difficult to implement in practice on a MAC grid, meaning that parallelization in this direction is undesirable.  A second advantage of this approach is that it enables the user to introduce different physical effects into the code concerning the interaction between the fluid and the solid wall (e.g. surface roughness), without interacting with the MPI aspects of the code.  In order to pass information between the sub-domains (which is necessary in finite-difference codes), the MPI library is used for straightforward synchronouse halo-swaps.

The research-level version of the TPLS code also uses calls from the Fortran code to the PETSc linear-algebra library to solve the Poisson equation for the pressure-correction step, which leads to a highly efficient parallelized pressure solver~\cite{scott2013}.
However, in order to provide students with a simple, robust and highly portable code  in the present version we have implemented instead the recently-discovered Scheduled Relaxation Jacobi (SRJ) scheme~\cite{Yang}. This is discussed in more detail below in Section~\ref{subsec:srj}, as it serves as a useful pedagogical digression concerning iterative methods for sparse linear equations.  In this way, the simplified code contains no calls to external libraries (other than standard MPI), and can be compiled on a desktop, a cluster, or a supercomputer using a standard Fortran 90 compiler with an MPI wrapper.

The code periodically dumps the pressure, velocities, and viscosity to a series of files for postprocessing and visualization.  In S-TPLS, this is done in a simple way, whereby all data is gathered to a single CPU processor and written to a single series of files.  The files are ASCII-formatted and configured for visualization using proprietary software (Tecplot).  However, because of the simplicity of the file structure the data can easily be accessed for visualization and postprocessing with other software.  A drawback of this approach is that the data output is performed on a single CPU processor, which is a bottleneck and limits the code's scalability.  For this reason, the full research-level version of TPLS uses parallel I/O with NetCDF~\cite{bethune2015}.

\subsection{Scheduled Relaxation Jacobi method}
\label{subsec:srj}

Iterative methods to solve Poisson's equation $\nabla ^2 \phi =  s$ numerically are often discussed in undergraduate computational science classes~\cite{garcia2000}. The Jacobi method, considered as the most primitive, in two-dimensions on a uniform rectangular grid of unit size takes the form (with $s=0$ for simplicity)
\begin{subequations}
\begin{equation}
\phi^{n+1}_{i,j} = \frac{1}{4} \left(\phi^{n}_{i+1,j}+\phi^{n}_{i-1,j}+\phi^{n}_{i,j+1}+\phi^{n}_{i,j-1}\right),
\end{equation}
where $n$ is the iteration index and $i,j$ indicate the $x$ and $y$ positions, respectively. Introducing the fixed relaxation parameter $\omega$ leads to the Jacobi with SOR scheme
\begin{equation}
\phi^{n+1}_{i,j} = \left(1-\omega\right)\phi_{i,j}^{n} + \frac{\omega}{4} \left(\phi^{n}_{i+1,j}+\phi^{n}_{i-1,j}+\phi^{n}_{i,j+1}+\phi^{n}_{i,j-1}\right).
\label{eq: Jacobi SOR}%
\end{equation}%
\end{subequations}
However, over-relaxation of the Jacobi method does not ensure convergence for all wavenumbers~\cite{Yang}. Hence this method is usually abandoned in favour of the Gauss-Seidel with SOR method which converges for $\omega \leq 2$.

The SRJ method, on the other hand, proposes the use of a combination of over-relaxation parameters ($\omega>1$) and under-relaxation parameters ($\omega<1$) applied to the Jacobi-SOR method as given by Equation~\eqref{eq: Jacobi SOR}. Essentially, a fixed number $M$ iterations of the Jacobi-SOR method are carried out, each of which has a prescribed relaxation parameter $\omega_{k}$ for $k$ ranging from $1$ to $P, P \geq 2$. These values are not necessarily unique, with each $\omega_{k}$ repeated $q_k$ times respectively, and the $M$-iteration cycle is then repeated until convergence. This method has been motivated by the fact that over-relaxation of the Jacobi method reduces the low wavenumber error while under-relaxation tends to reduce the high-wavenumber error~\cite{Yang}.

In the context of this report, the SRJ method for a three-dimensional problem, whose form is analogous to Equation~\eqref{eq: Jacobi SOR}, has been incorporated into \simplesend. For given values of $(N_x,N_y,N_z)$ on a unit domain, one can choose an appropriate set of relaxation parameters $w_k$ based upon a characteristic number of grid points $N$. The relaxation schedule (i.e. the order in which the $w_k$ should appear in the $M$-iteration) is then obtained courtesy of the Matlab script provided in Reference~\cite{Yang}.

\section{Validation}
\label{sec:validation}

Given the sheer size of the \simples code, its combination of many algorithms, and its incorporation of the MPI parallel programming methodology, it is essential to be able to validate its accuracy and fidelity to the underlying equations of motion, before running expensive simulations and making predictions concerning the properties of the flow under consideration.

Linear stability analysis provides a rigorous test case with which to validate \simples -- at least prior to incorporation of the LES technique.
The reason is that the linear stability analysis of a steady base state yields  a non-trivial  quasi-analytical temporally-evolving perturbed state, which can be compared to a similar state emanating from a  direct numerical simulation.
In the context of the present model problem, the linear stability theory is known as an Orr--Sommerfeld analysis~\cite{orr}.  While it is not the purpose of this paper to present Orr--Sommerfeld analysis as a pedagogical tool, we present it herein as a best-practice for validation.  More elaborate versions of the same linear theory can be used to validate research-level problems, for example in two-phase flow~\cite{lennon_1,valluri2010linear}.

A further justification for validating the code prior to the incorporation of the LES technique is that the latter is a simple `add on' to the basic hydrodynamics modules in the code.  Thus, the requirement that the pre-LES version of the code should agree precisely with Orr--Sommerfeld theory is a necessary condition for the code's correctness.
Subsequent to the incorporation of the LES technique, validation can be done \textit{a posteriori}, e.g. by examining the properties of the numerically computed mean flow and comparing to previous numerical and experimental works.  This second aspect is discussed in Section~\ref{sec:results}.
For these reasons, the Orr--Sommerfeld theory is introduced and then used in this section as a stringent test to demonstrate the validity of the \simples code.

\subsection{Orr--Sommerfeld analysis}

 We consider the laminar flow analogue to the model problem given in Section~\ref{sec:problem}. In this case, the problem reduces to a two-dimensional one i.e. in the $xz$ plane, with the flow being both uni-directional and steady. As such, enforcing no-slip boundary conditions on the walls leads to the standard Poiseuille velocity profile
\begin{equation}
U_0\left(z\right) = \frac{h^2}{2\mu}\left|\frac{dp}{dx}\right|\frac{z}{h}\left(1-\frac{z}{h}\right),
\end{equation}
where we make use of the fact that the pressure gradient is inherently negative and $L_{z}=h$ is the channel height. Using the same non-dimensional formulation as in Section~\ref{sec:problem}, this can be written in non-dimensional form as
\begin{equation}
U_0 \left(z\right)=Re_* z\left(1-z\right).
\label{eq:Poiseuille non-d}
\end{equation}
The idea behind the Orr--Sommerfeld analysis is to introduce a small perturbation to this base state profile (denoted using subscripts $0$) of the form
\begin{equation}
\left(u,w,p\right)=\left(U_0+\delta u,\delta w,p_0+\delta p\right).
\end{equation}
We now perform a linear stability analysis by substituting this velocity profile into the non-dimensional NS equations~\eqref{eq:NS_all} and ignoring terms which are quadratic in the perturbation velocities. Linearising about the base flow yields the following set of equations
\begin{subequations}
\begin{eqnarray}
\frac{\partial \left(\delta u\right)}{\partial t} + U_0\frac{\partial \left(\delta u\right)}{\delta x} +\delta w \frac{\partial U_0}{\partial z} = -\frac{\partial \left(\delta p\right)}{\partial x} + \frac{1}{Re_*}\nabla^2 \delta u, \label{eq:OS_xp} \\
\frac{\partial \left(\delta w\right)}{\partial t} + U_0 \frac{\partial \left(\delta w\right)}{\partial x} = - \frac{\partial \left(\delta p\right)}{\partial z}+ \frac{1}{Re_*}\nabla^2 \delta w,\label{eq:OS_zp} \\
\frac{\partial \left(\delta u\right)}{\partial x} + \frac{\partial \left(\delta w\right)}{\partial z} = 0. \label{eq:OS_cont}
\end{eqnarray}%
\end{subequations}%
As the flow is two-dimensional we can relate the perturbation velocities to the streamfunction $\Psi\left(x,z,t\right)$ via
\begin{equation*}
\delta u = \frac{\partial \Psi}{\partial z}, \quad \delta w = -\frac{\partial \Psi}{\partial x};
\end{equation*}
moreover, we can make a normal-mode decomposition $\Psi = \exp\left[\imag \alpha\left(x-ct\right)\right] \psi\left(z\right)$, where $\alpha \in \mathbb{R}$ and $c \in \mathbb{C}$ are the wavenumber and speed of the disturbance, respectively. After a number of steps, one arrives at the Orr--Sommerfeld equation
\begin{equation}
\imag \alpha\left(U_{0}-c\right)\left(\frac{\partial^2}{\partial z^2}-\alpha^2\right)\psi -\imag \alpha U_{0}'' \psi = \frac{1}{Re_*}\left(\frac{\partial^2}{\partial z^2} - \alpha^2 \right)^2\psi \label{eq: OS Equation}
\end{equation}
which, supplemented with the no-slip boundary conditions $\psi\left(z\right)=\psi'\left(z\right)$ for $z=0,1$, presents an eigenvalue problem for $c$. For our purposes, it is important to note that since $\Psi \propto \exp\left[\imag \alpha\left(x-ct\right)\right] = \mathe^{\imag \alpha x}\mathe^{\lambda t}$, where $\lambda = -\imag \alpha c$, the perturbation velocities will grow exponentially in time if $\Re\left(\lambda\right)>0$ (note: since $\lambda=-\imag\alpha c$, exponential growth with $\Re(\lambda)>0$ corresponds to $\Im(c)>0$ also).
%
%
%
 Moreover, the perturbation velocities also undergo an oscillation with period $T = 2 \pi / \omega = 2 \pi / \left| \Im\left(\lambda\right) \right|$.

The foregoing description pertains to all Reynolds numbers, yet a more precise classification of the solution to the eigenvalue problem~\eqref{eq: OS Equation} is provided
in the  context of hydrodynamic instability by introducing the so-called critical Reynolds number $Re_{*\mathrm{c}}$:
\begin{itemize}
\item For $Re_*> Re_{*\mathrm{c}}$ one can find a range of unstable wavenumbers for which $\Re(\lambda)>0$ and the perturbations grow and contaminate the base state.  This is referred to in the literature as a \textit{supercritical} case~\cite{schmid2012stability}.
\item For $Re_*<Re_{*\mathrm{c}}$ we have $\Re(\lambda)<0$ for all wavenumbers, hence all perturbations of sufficiently small initial amplitude die out as $t\rightarrow\infty$.  This is referred to in the literature as a \textit{subcritical} case~\cite{schmid2012stability}.
\end{itemize}
%
%
The critical Reynolds number for Poiseuille flow was first obtained in Reference~\cite{Orszag}
and is known to be $Re_{*\mathrm{c}}\approx 214.9$.
(Note that $Re_{*\mathrm{c}}$ is given here in terms of the friction velocity and the channel height, yet in Reference~\cite{Orszag} the critical Reynolds number is given in terms of  the mean velocity and the channel half-height -- under this rescaling the critical Reynolds number for Poiseuille flow assumes the approximate value $5772$.)

\subsection{Simulation results}
The \simples code is modified in order to perform an Orr--Sommerfeld analysis. Simulation parameters used and predicted values for $\Re\left(\lambda\right)$ and $T$ are given in Table~\ref{table:OS_sim_par}.  In particular, the Reynolds number is taken to be $Re_*=500$, and the wavenumber is taken to be $\alpha=2\pi/4$, corresponding to a supercritical case with $\Re\left(\lambda\right)>0$.  The eigenvalue analysis is conducted using an in-house Orr--Sommerfeld solver which has been validated carefully against the results of
Orszag~\cite{Orszag}.  This is compared with a simulation in a channel of length $L_x=4$, chosen so as to correspond to the wavelength of the unstable mode.

\begin{table}
\begin{tabular}{|c|c|c|c|c|c|c|c|c|c|c|}
\hline
$Re_*$&$N_x$&$N_y$&$N_z$&$\Delta x$ & $\Delta y$ & $\Delta z$ &$\Delta t$& $\Re\left(\lambda\right)$& $\Im\left(\lambda\right)$ & $T$\\
\hline
\hline
500&1200&3&300& $\left(300\right)^{-1}$ & $\left(300\right)^{-1}$ & $\left(300\right)^{-1}$ &$10^{-5}$&1.8056&-33.873&0.185\\
\hline
\end{tabular}
\caption{Simulation parameters used for Orr--Sommerfeld analysis of the \simples code. Note the small number of grid points used in the $y$ direction as the problem is two-dimensional in nature.}
\label{table:OS_sim_par}
\end{table}

\begin{figure}[htb]
\centering
\subfigure[]{\includegraphics[width=0.45\textwidth]{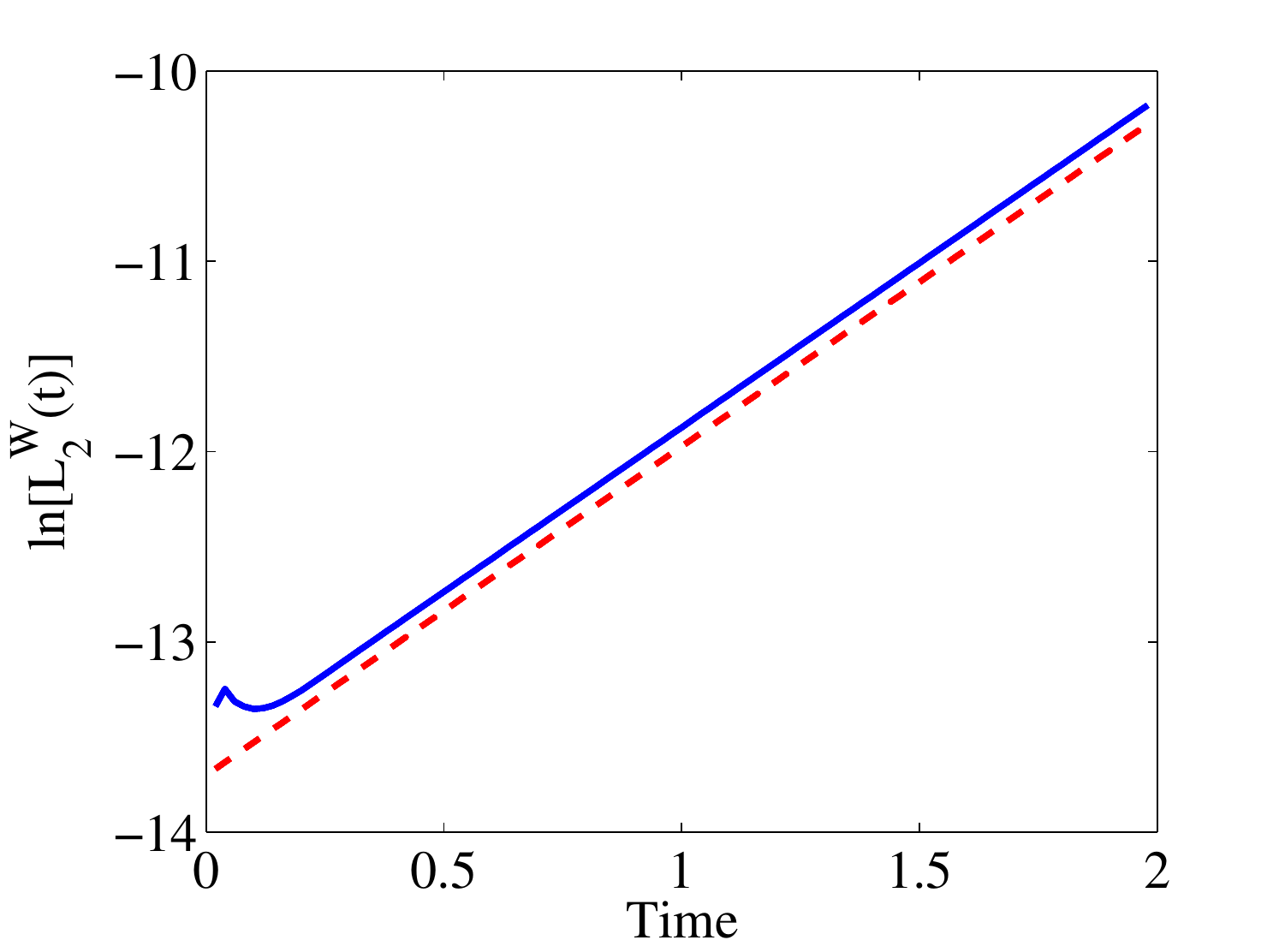}}
\subfigure[]{\includegraphics[width=0.45\textwidth]{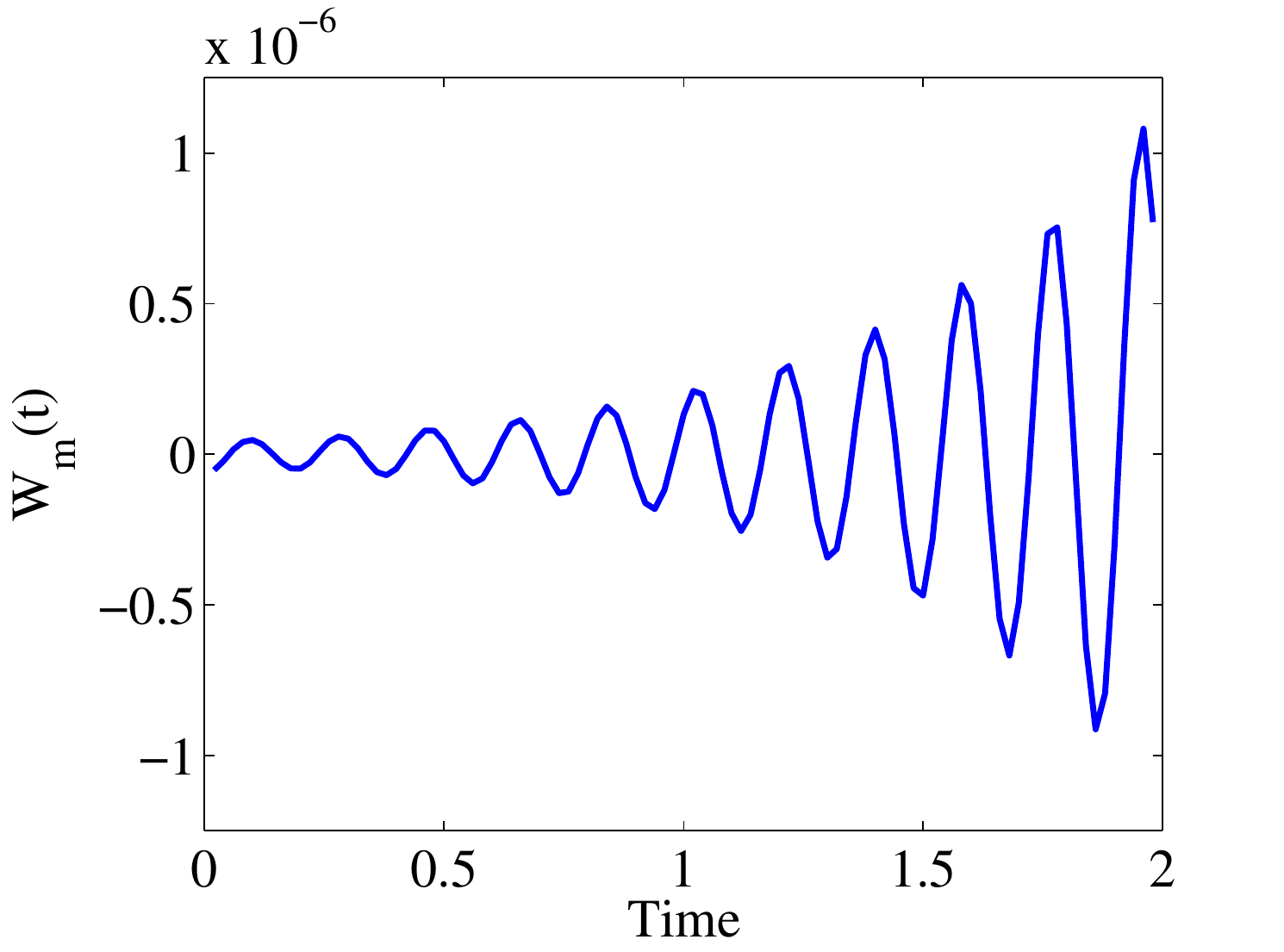}}
\subfigure[]{\includegraphics[width=0.45\textwidth]{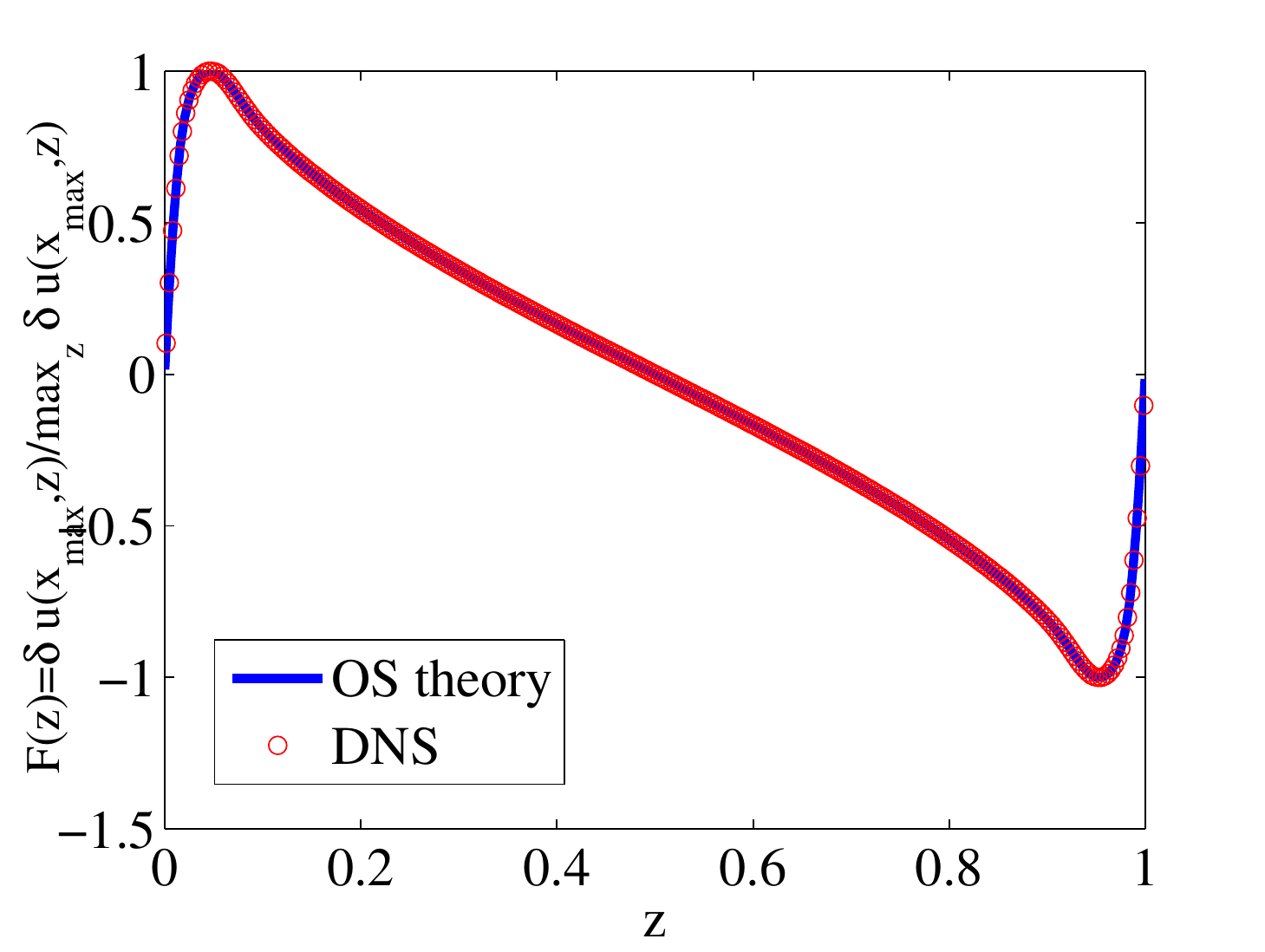}}
\caption{(a) Plot of $\ln\left[L_{2}^{w}\left(t\right)\right]$ with solid lines indicating the numerical result and dashed lines the linear fit corresponding to $\lambda \approx 1.7293$; (b) plot of $W_{m}\left(t\right)$; (c) plot of $F(z)$ showing the agreement between the DNS and the OS analysis for the spatial structure of the perturbations.}
\label{fig:OS_plots}
\end{figure}

The initial condition needed to trigger the numerical instability is introduced via an initial incompressible perturbation velocity field added to the base-state profile given by  Equation~\eqref{eq:Poiseuille non-d}.   The perturbation velocity field is given here:
\begin{subequations}
\begin{equation}
\delta u = - A\left(x\right) \frac{d f\left(z\right)}{dz}, \quad \delta w = f(z) \frac{d A\left(x\right)}{dx},
\label{eq: perturbation velocity numerical}
\end{equation}
where $f\left(z\right)$ and $A\left(x\right)$ are defined as
\begin{eqnarray}
A\left(x\right) = \epsilon\left(C \cos\left(\alpha x\right) + S \sin\left(\alpha x\right) \right), \quad f\left(z\right) = \tfrac{1}{2} + \tfrac{1}{2}\sin\left(\frac{2\pi z}{L_{z}}-\frac{\pi}{2}\right).
\label{eq:f_def}
\end{eqnarray}%
\end{subequations}%
Here $\epsilon$ is a small value (taken to be $10^{-4}$ in the code) while $C$ and $S$ are constants between $0$ and $1$ (taken to be $\pm \left(2\right)^{-1/2}$, respectively). The predictions of the eigenvalue analysis can then be tested against the numerical results by considering the following quantities:
\begin{subequations}
\begin{eqnarray}
 L_{2}^{w}(t) &=& \left|\left| \delta w\left(x,z,t\right) \right|\right|_{2} = \left[\int_{0}^{L_x}\int_{0}^{L_z} \left(\delta w\right)^2 dz dx \right]^{1/2},\\
 W_{m}(t) &=& \delta w\left(\frac{L_x}{2},\frac{L_z}{2},t\right),\\
F(z)&=&\frac{\delta u(x_0,z,t)}{\max_{z\in[0,1]} \delta u(x_0,z,t)},
\end{eqnarray}
\end{subequations}
where $x_0$ corresponds to the location of the global maximum of $\delta u(x,z,t)$ in the whole flow domain.
Theoretically, the $L_{2}$ norm of the $\delta w$ perturbation ($L_2^{w}\left(t\right)$) should grow exponentially with time with a rate given by $\Re\left(\lambda\right)$,  $W_{m} \left(t\right)$ should be an oscillating function with an exponentially-growing envelope and period given by $T$ and finally, $F(z)$ should be time-independent.  The numerical results obtained are plotted in Figure~\ref{fig:OS_plots}.
The predicted linear growth in $\ln\left(L_2^{w}\left(t\right)\right)$ is evident from figure \ref{fig:OS_plots}(a), where the growth rate is found to be $\lambda_{S} = 1.7293$. This agrees well with the predicted value (the two differing by an error of $\approx 4\%$) and decreased error is expected upon further grid refinement. Furthermore, the predicted behaviour of $W_{m}\left(t\right)$ is also observed in Figure~\ref{fig:OS_plots}(b). The period of the oscillation is found to be $T_{S} \approx 0.186 \pm 0.004$, in excellent agreement with the predicted value of $T=0.185$.  The spatial structure of the flow is encoded in the function $F(z)$; excellent agreement in $F(z)$  between the numerical simulation and the theoretical calculation arising from the Orr--Sommerfeld eigenvalue analysis is evidenced in Figure~\ref{fig:OS_plots}(c).

\section{Performance Analysis}
\label{sec:performance}

Before embarking on a large-scale LES of turbulent channel flow, it is of interest to know the most efficient way in which to deploy the parallelism of \simples with both the SRJ pressure solver and the LES technique incorporated into the code.  In this section, we therefore
investigate the strong scaling behaviour of the code i.e. how its performance varies with the number of cores used for a fixed problem size.  This information then enables us to choose the number of CPU cores so as to maximize the efficiency of the code.  The information is also of more general interest, as it enables one to quantify the trade-off between execution time and computational resources.

This performance analysis is straightforward: using an appropriate domain and grid resolution, the \simples solver is run for a total of $1000$ iterations using $N_{p}$ CPU cores. Omitting time associated with initialisation and periodic file output, the execution time of the code is recorded as $T_{p}$ and the number of processes used varied over the range $N_{p}=24\mbox{--}1008$.  The value $N_p=24$ is a result of the architecture of the machine used for the calculation (Fionn, ICHEC): each compute node on Fionn consists of 24 cores, and an efficient use of the resources requires that an integer number of nodes be reserved for each simulation.
 The analysis parameters are outlined in Table~\ref{table:PA_sim_par}. One notes the use of a `long' domain and a higher resolution in the wall-normal direction to that used in the streamwise and spanwise directions. This is due to the fact that the most vigorous turbulence production occurs close to the walls~\cite{Pope}, thus requiring a higher grid resolution.


\begin{table}[b]
\begin{tabular}{|c|c|c|c|c|c|c|c|c|}
\hline
$Re_*$&$N_x$&$N_y$&$N_z$& $N_{T}$& $\Delta x$ & $\Delta y$ & $\Delta z$ &$\Delta t$\\
\hline
\hline
360&288&126&120& 4,354,560 & $\left(36\right)^{-1}$ & $\left(36\right)^{-1}$ & $\left(120\right)^{-1}$ &$5 \times 10^{-5}$ \\
\hline
\end{tabular}
\caption{Parameters used for the performance analysis of the \simples code incorporating both the SRJ solver and LES technique.}
\label{table:PA_sim_par}
\end{table}

\begin{figure}
\centering
\subfigure[]{\includegraphics[width=0.4\textwidth]{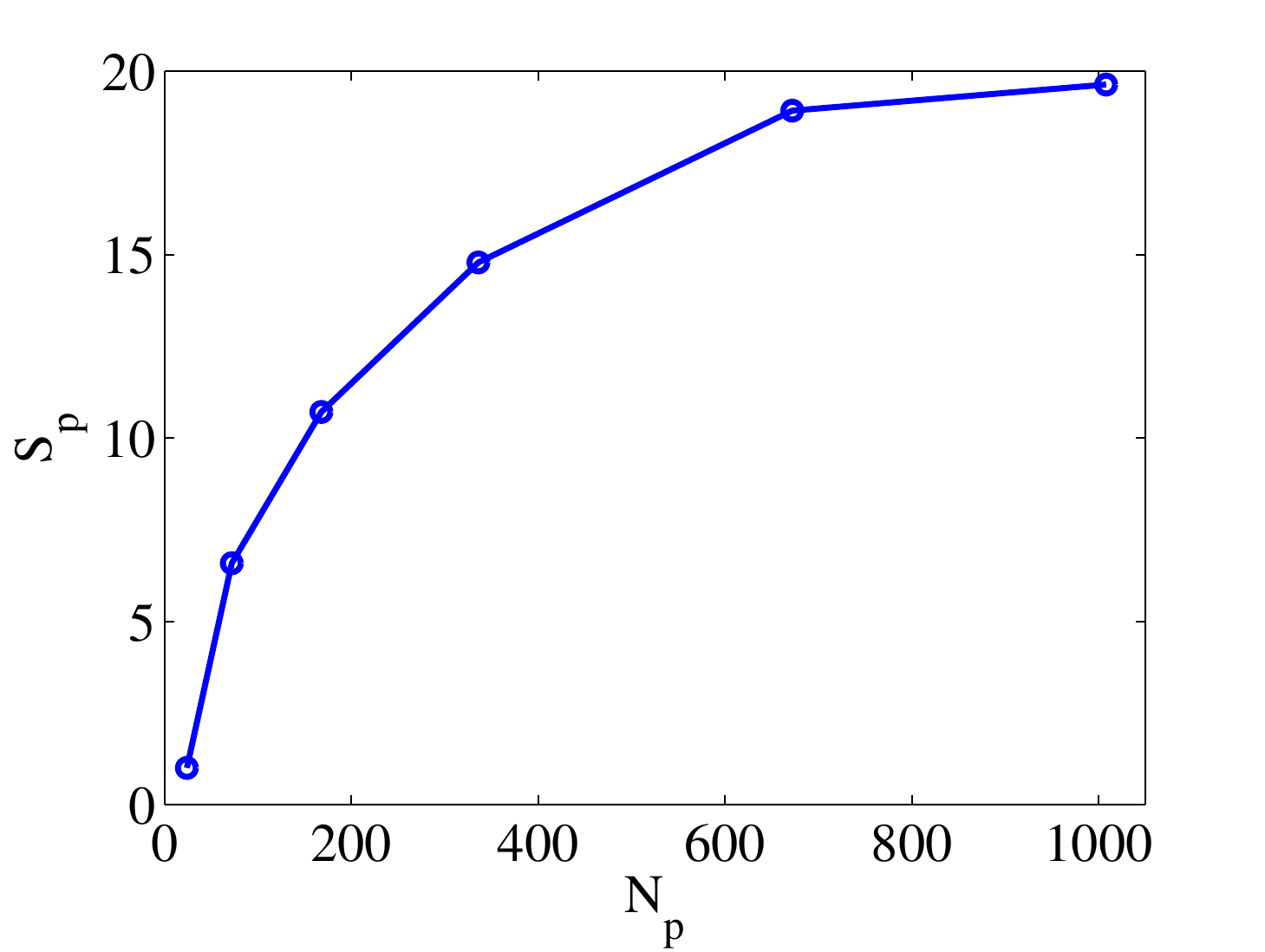}}
\subfigure[]{\includegraphics[width=0.4\textwidth]{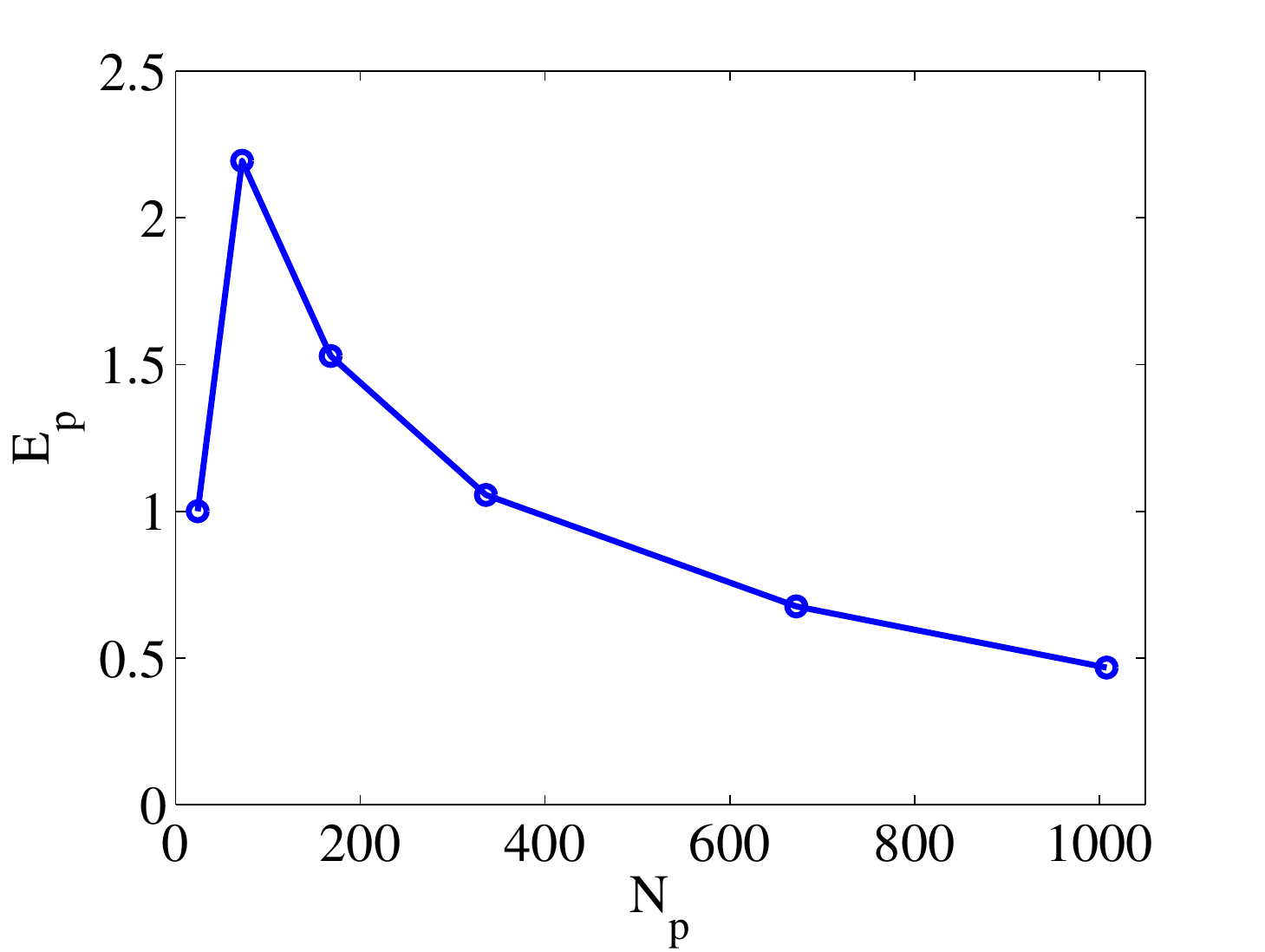}}
\caption{(a) Speedup curve $S_{p}$ and (b) parallel efficiency curve $E_{p}$.}
\label{fig:speed_up}
\end{figure}

Having obtained the execution times (given in Table \ref{table:Exec_times}), one can calculate associated quantities such as the speedup $S_{p}$ and parallel efficiency $E_{p}$ which are given by
\begin{equation}
S_{p}= \frac{T_{24}}{T_{p}}, \quad E_{p} = \frac{24 T_{24}}{N_{p} T_{p}}.
\end{equation}
These quantities are plotted in Figure~\ref{fig:speed_up}. Evidently, one can observe from Figure~\ref{fig:speed_up}(a) that super-linear speedup is achieved for process counts $N_{p} \leq 336$, with sub-linear speedup beyond this point. Moreover, one can observe that $E_{p} < 1$ for values of $N_{p} \gtrsim 385$.  Superlinear speedup is generally associated with an efficient use of the cache by the code as memory required by the sub-problem on a particular CPU core is reduced under the domain decomposition.  In contrast, sublinear speedup at higher process numbers is associated with communication overheads, which we investigate below.

To understand the communication overhead associated with running the code in parallel, the code was profiled using the Allinea MAP tool provided on the ICHEC system. Profiling the code allows one to investigate how much time is spent on each individual part of the code. In parallel computing, profiling is extremely useful as it identifies the most time-intensive parts of the code which can then be refined and optimized, so as to achieve increased efficiency. Using the same parameters as outlined in Table~\ref{table:PA_sim_par}, two sample profiles are constructed (Table~\ref{table:Code_profile}).
\begin{table}[t]
\begin{tabular}{|c|c|}
\hline
Number of processors $N_{p}$& Execution time $T_{p}$ (seconds)\\
\hline
\hline
24 & 3669.86 \\
$3 \times 24 = 72$ & 557.45 \\
$7 \times 24 = 168$ & 342.76 \\
$14 \times 24 = 336$ & 248.26 \\
$28 \times 24 = 672$ & 193.94 \\
$42 \times 24 = 1008$ & 186.91 \\
\hline
\end{tabular}
\caption{Execution times $T_{p}$. Note that the choice of multiples of $24$ for the number of processes $N_{p}$ was due to the architecture of the machine used (Fionn, ICHEC), and is not a requirement of the code itself.}
\label{table:Exec_times} 
\end{table}
\begin{table}[b]
\begin{tabular}{|c|c|c|c|c|}
\hline
Number of processors $N_{p}$ & Pressure calculation & MPI send-receives & Velocity calculation & Other \\
\hline
\hline
72& 51.3\% & 39.6\% & 8.7\% & 0.4\% \\
\hline
\end{tabular}
\caption{Profile illustrating the most time intensive parts of the single phase SRJ LES solver. Note that MPI send-receives represents time spent by processors communicating with one another.  The percentage sign refers to the percentage of total time.}
\label{table:Code_profile}
\end{table}
One can see that the solution of the Poisson equation for the pressure  is by far the most intensive aspect of the `maths' part of the code, far outweighing the time spent computing the velocity field. This is understandable given that one must use an iterative process for a very weakly diagonally-dominant system (i.e. Poisson's equation) in order to solve for the pressure. Communication overheads (MPI send-receives) are also a dominant feature, with over half the execution time ($55.7\%$) spent on communication between processors for $N_{p}=216$. These overheads will continue to increase as $N_{p}$ is increased, leading to a decrease in $E_{p}$ as evident from Figure~\ref{fig:speed_up}(b).

In summary, the efficiency of the parallel code is much reduced at large CPU core counts.  This can be viewed as the penalty for using \simplesend, as the parallel efficiency of the research-level code which uses the PETSc and NetCDF libraries has been measured to be $0.9$ at $N_p>2000$~\cite{scott2013}.  For the present pedagogical applications, this loss of efficiency is not severe, and only manifests itself at large CPU core counts (which would anyway correspond to research-level simulations).  Also, the loss of efficiency is compensated by the increase in the code simplicity and portability.

\section{Results of the LES}
\label{sec:results}

It is well known that the supercritical instability of channel Poiseuille flow is a route to turbulence~\cite{drazin2004hydrodynamic}.  Subcritical routes also exist, and are a matter of intense interest in the literature~\cite{schmid2012stability}.  There is therefore a nice connection between the direct numerical simulation of the unstable channel Poiseuille flow in Section~\ref{sec:validation} and the study of fully developed turbulence at supercritical Reynolds numbers.  In the present section we study the latter, using the large-eddy concept developed in Sections~\ref{sec:intro}--\ref{sec:problem}.  Rather than introduce new research, the aim here is to showcase the ability of \simples to carry out intensive simulations in an efficient manner, and the computational results are compared with existing standard results from the literature.  In presenting the below results, we omit the overbar over the spatially-filtered velocity and pressure fields, since it is clear that we are dealing with an LES simulation.

We focus on fully developed turbulence, rather than the laminar-turbulent transition.  Therefore, the simulation is seeded with a turbulent-like initial condition, so that the simulation settles down rapidly to a fully-developed state, for which the statistics of the flow can be gathered.  Specifically, we take~\cite{batten2004interfacing,benocci2006large}
\begin{subequations}
\begin{equation}
\vecu(\vecx,t=0)=\silly  U_0(z)+\nabla\times\left[f(z)\vecA(\vecx)\right],\qquad
p(\vecx,t=0)=\frac{dp}{dx}x,
\end{equation}
where  $U_0(z)$ is the base-state Poiseuille flow given by Equation~\eqref{eq:Poiseuille non-d}, the function $f(z)$ is given by Equation~\eqref{eq:f_def}, and $\vecA(\vecx)$ is a collection of random Fourier modes given by
\begin{equation}
A_{i}(\vecx)= \sum_{k_x=1}^{C_{k_x}} \sum_{k_y=1}^{C_{k_y}} \sum_{k_z=1}^{C_{k_z}} \left[ S_{i}\left(k_x,k_y,k_z\right) \sin(\veck\cdot\vecx) + C_{i}\left(k_x,k_y,k_z\right) \cos(\veck\cdot\vecx) \right],
\label{eq:ai_def}
\end{equation}
where we have a sum over the wavenumber $\veck$:
\begin{equation}
\veck = \left(\frac{2\pi}{L_x}k_x,\frac{2\pi}{L_y}k_y,\frac{2\pi}{L_z}k_z\right),
\end{equation}%
\end{subequations}%
and the constants $S_{i}\left(k_x,k_y,k_z\right)$ and $C_{i}\left(k_x,k_y,k_z\right)$ are random numbers between $0$ and $1$.  The term $\nabla\times\left[f(z)\vecA\right]$ represents a manifestly incompressible fluctuation, while $\silly U_0(z)$ represents a crude approximation to the turbulent mean flow, which is reduced and `flattened' with respect to its laminar counterpart~\cite{Pope} -- we therefore take $\silly=1/3$.

\begin{figure}
	\centering
		\subfigure[]{\includegraphics[width=0.45\textwidth]{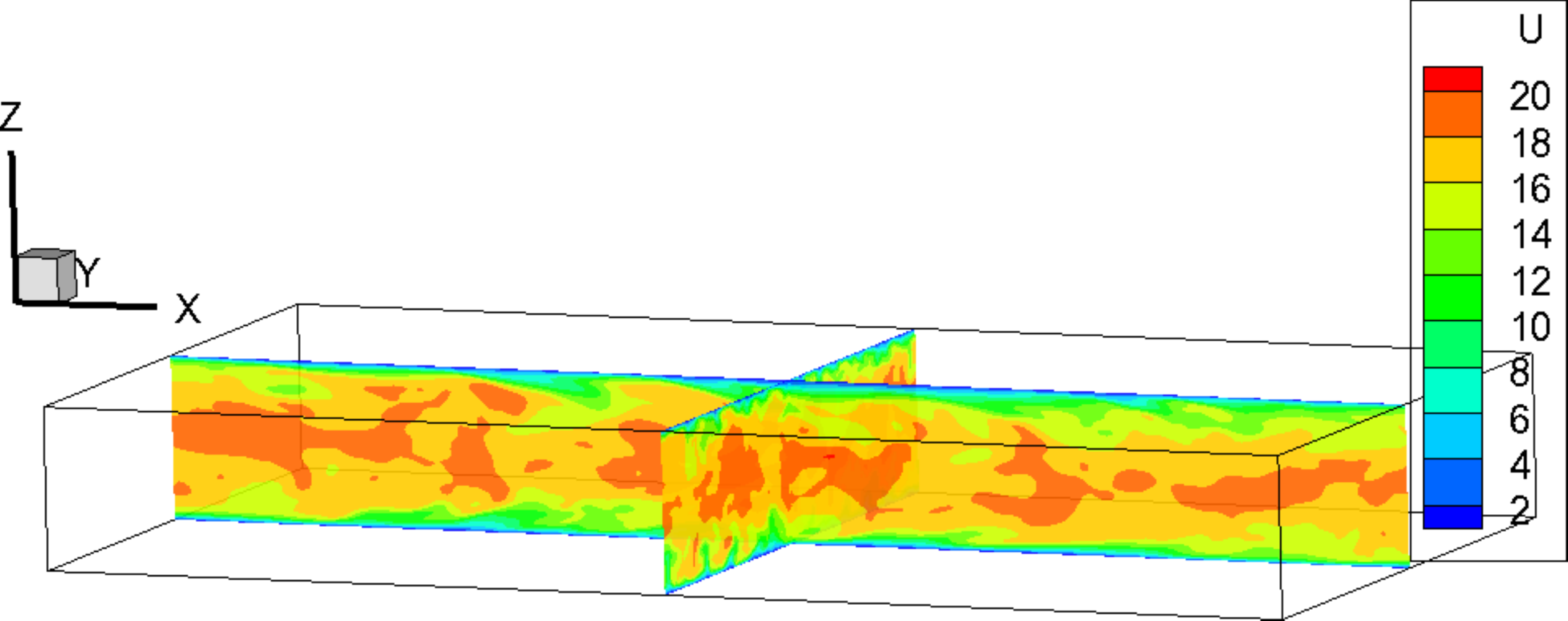}}
		\subfigure[]{\includegraphics[width=0.45\textwidth]{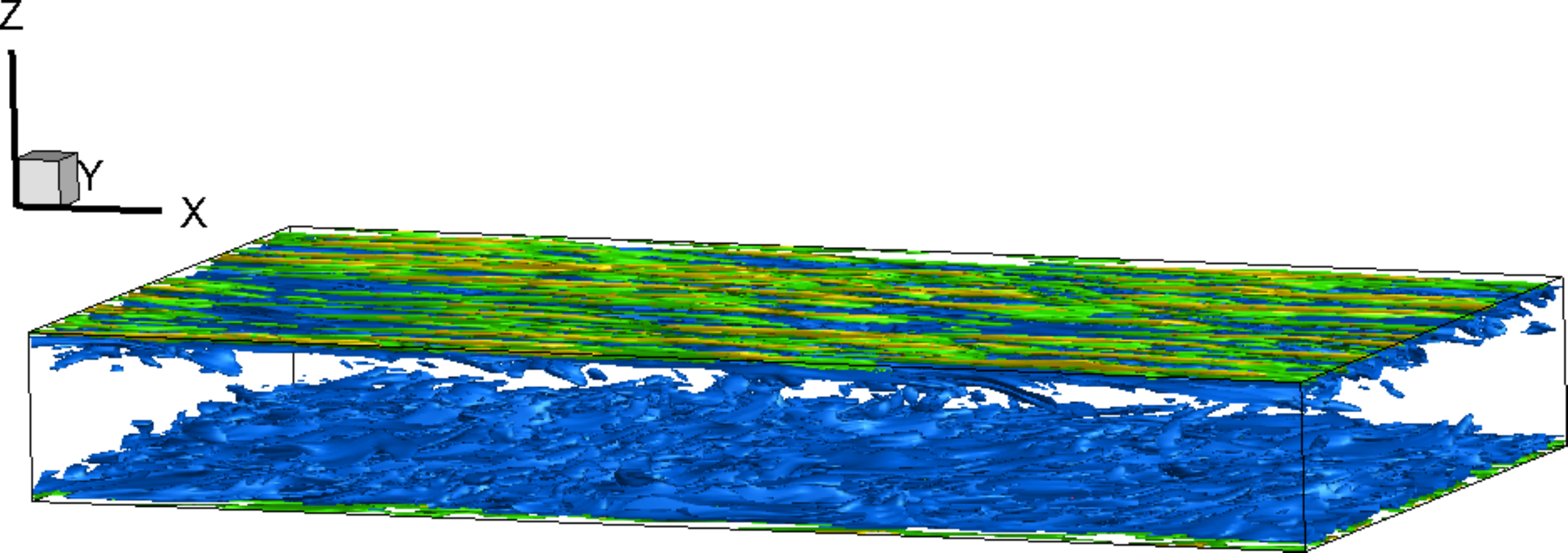}}
		\caption{Snapshot of instantaneous motion, $t=21.6$.  Panel (a): Instantaneous streamwise velocity showing the maximum streamwise velocity near the centreline; Panel (b): isosurfaces of instantaneous vorticity magnitude showing intense turbulence generation near the walls.  The isosurfaces in green closest to the walls correspond to $|\nabla\times\vecu|=300$; the isosurfaces in blue with contributions further from the walls correspond to $|\nabla\times\vecu|=100$.}
	\label{fig:snapshot_turb}
\end{figure}

\begin{figure}
	\centering
		\includegraphics[width=0.5\textwidth]{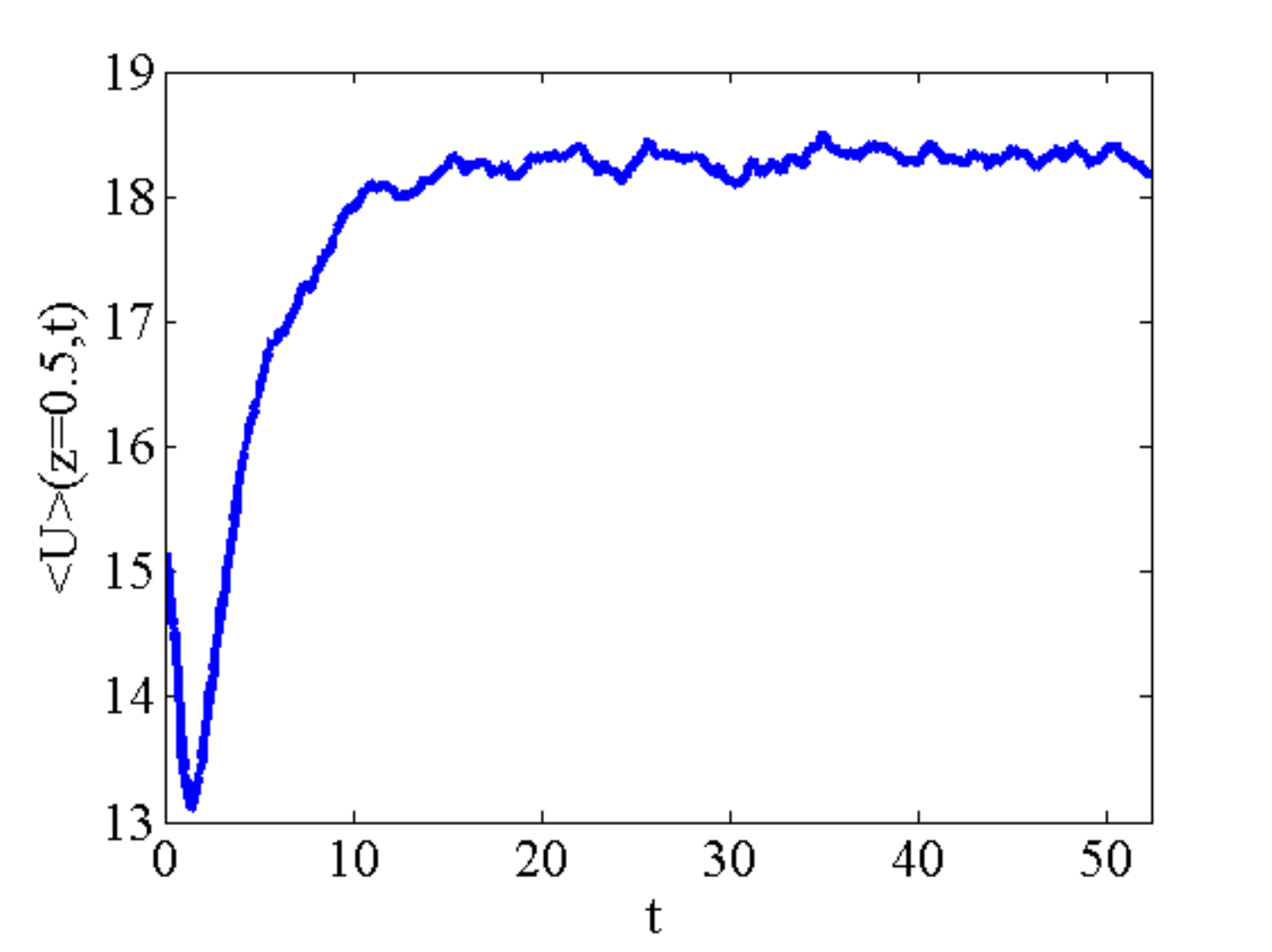}
		\caption{Time dependence of the mean centreline velocity showing the convergence to a statistically steady state.}
	\label{fig:u_cl}
\end{figure}

Based on the above, an LES was performed using the parameters in Table~\ref{tab:LES_sim_par}, using five Fourier modes in each spatial direction for the initial condition in Equation~\eqref{eq:ai_def}.
\begin{table}
\begin{tabular}{|c|c|c|c|c|c|c|c|}
\hline
$Re_*$&$N_x$&$N_y$&$N_z$&$L_x$ & $L_y$ & $L_z$ &$\Delta t$ \\
\hline
\hline
360&240&140&120& $8$ & $4$ & $1$ &$10^{-4}$\\
\hline
\end{tabular}
\caption{Simulation parameters used for LES.  The number of CPU cores used is $N_p= 336$ on Fionn, chosen to correspond to a regime wherein the parallel efficiency of the code is close to 100\%.  Also, the size of the grid in the $x$- and $y$-directions is chosen with respect to a reference paper~\cite{Kim} so that the turbulent structures do not interact with each other through the periodic boundary conditions.}
\label{tab:LES_sim_par}
\end{table}
Snapshot results are shown in Figure~\ref{fig:snapshot_turb}.
The snapshots reveal the expected qualitative features of turbulent channel flow, in particular the streamwise maximum velocity near the channel centreline and the intense turbulence generation near the walls, as evidenced by the large vorticity magnitude in Figure~\ref{fig:snapshot_turb}.
The code was run for $53$ dimensional time units, and a statistically steady state was attained after 20 time units.  The onset of the statistically steady state was determined by inspection of the average channel centreline velocity $L_x^{-1}L_y^{-1}\int_0^{L_x}\mathd x\int_0^{L_y}\mathd y \,{u}(x,y,z=0.5,t)$: after some transience this fluctuates around a steady value (Figure~\ref{fig:u_cl}).

\begin{figure}
	\centering
		\subfigure[]{\includegraphics[width=0.45\linewidth]{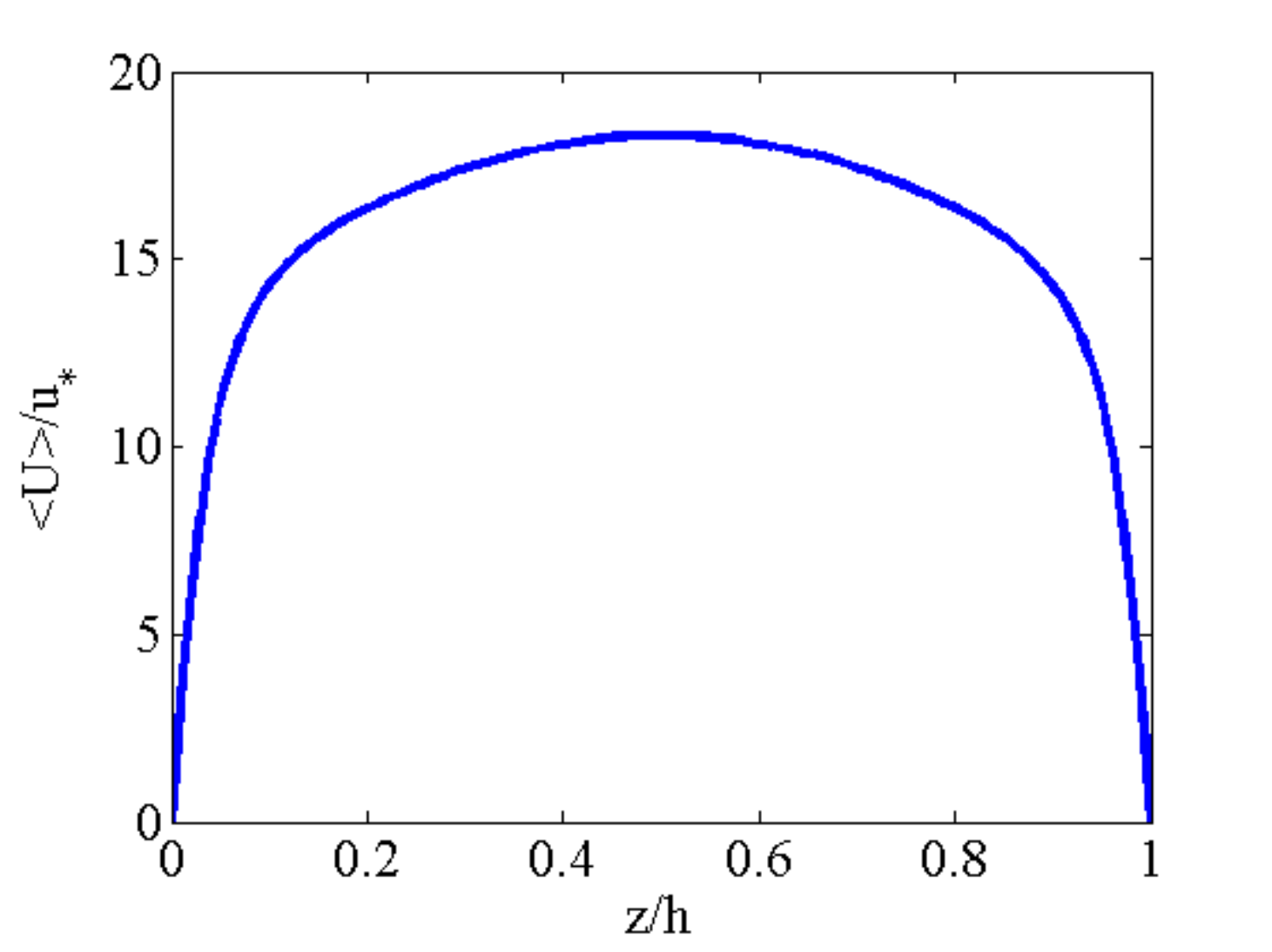}}
		\subfigure[]{\includegraphics[width=0.45\linewidth]{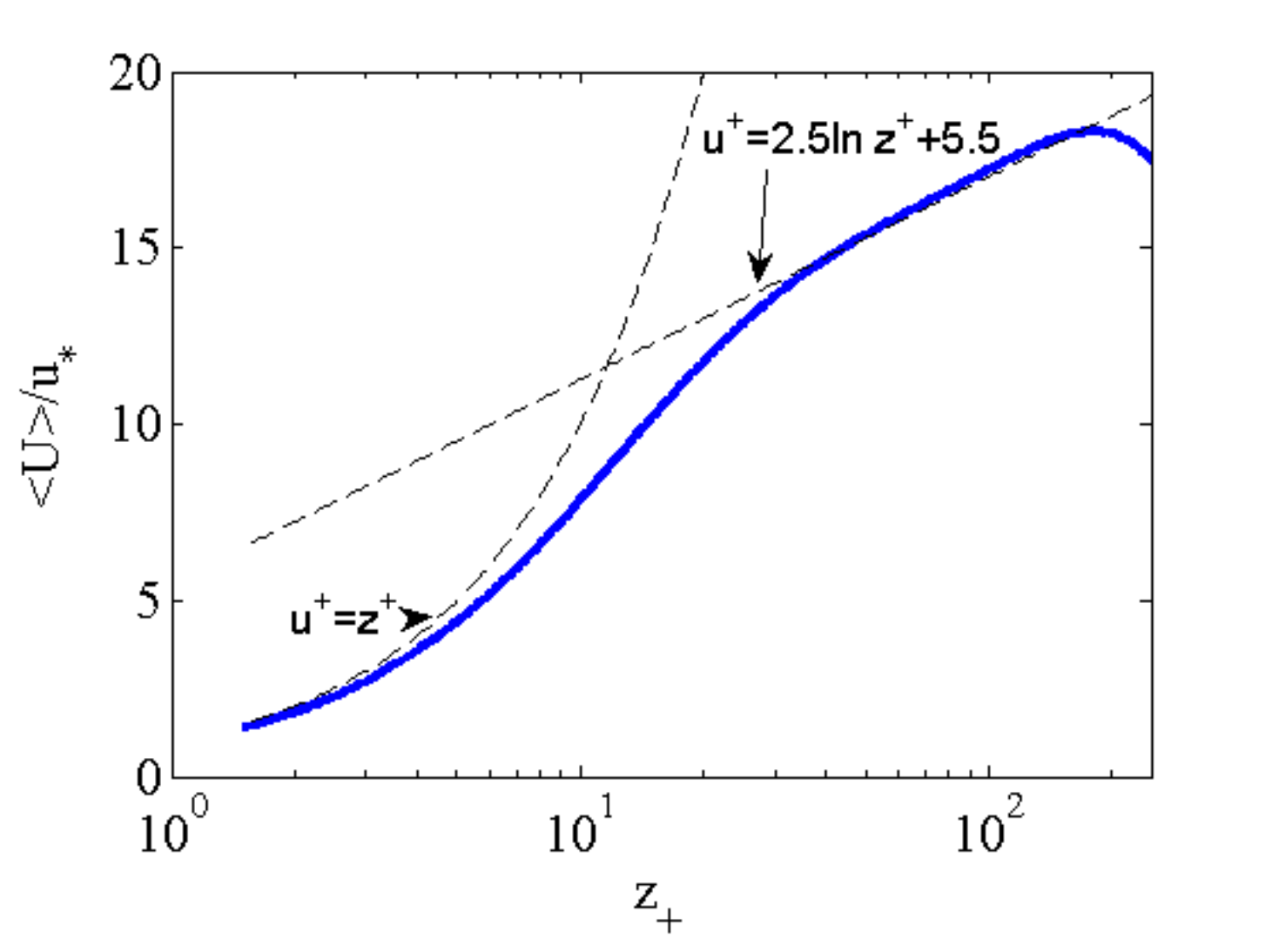}}
		\caption{The mean velocity profile for fully-developed flow at $Re_*=360$.  The centreline velocity agrees with the correlation $U_\mathrm{max}/u_*=(1/0.4)\log Re_*+3.47$ from the literature~\cite{Pope}.  (b) The mean velocity profile, showing the viscous sublayer $\langle u\rangle\sim z^+$ and $\langle u\rangle=(1/0.4)\log z^++5.5$: the logarithmic layer appears in the region $50<z^+<150$.  Here $z^+=zRe_*$ denotes the wall-normal coordinate in wall units.}
	\label{fig:u_mean}
\end{figure}

The instantaneous velocity fields (such as that in Figure~\ref{fig:snapshot_turb}(a)) are averaged over time $t\geq 20$ and over the $x$- and $y$-directions to produce space-time average quantities: for a property $\psi(x,y,z,t)$ we define the spacetime average as
\[
\langle \psi\rangle(z)=\frac{1}{t_2-t_1}\frac{1}{L_xL_y}\int_{t_1}^{t_2}\mathd t\int_0^{L_x}\mathd x\int_0^{L_y}\mathd y\,\psi(x,y,z,t),
\]
with $t_1=20$ and $t_2=53$.  The structure of the mean flow $\langle {u}\rangle(z)$ is obtained in this way and the results are presented in Figure~\ref{fig:u_mean}.
The characteristic `flattenend' profile is shown in Figure~\ref{fig:u_mean}(a).  The same profile is shown in wall units on a semi-lograithmic scale in Figure~\ref{fig:u_mean}(b).  This profile is consistent with the `universal' description of wall-bounded turbulence~\cite{lesieur2005large,Pope}, with $\langle u\rangle \sim z^+$ near the wall, and a `log layer' $\langle u\rangle=(1/0.4)\log z^++5.5$ in an intermediate zone between the wall and the channel centreline~\cite{lesieur2005large,Pope}.  Here $z^+=zRe_*$ denotes the wall-normal coordinate expressed in wall units.

We also compare these results with our knowledge of the `bulk' properties of the flow, i.e. properties obtained by averaging over all spatial dimensions.  We introduce
\[
U_{\mathrm{mean}}=\frac{1}{L_z}\int_0^{L_z}\mathd z\,\langle u\rangle.
\]
Based on Figure~\ref{fig:u_mean}, this is computed to be $U_\mathrm{mean}=15.39$, hence $Re_{\mathrm{mean}}=U_\mathrm{mean}H/\nu=5542$.  The centreline velocity derived from the same figure is $18.30$, in agreement with the known correlation~\cite{Pope} $U_\mathrm{max}\approx (1/0.4)\log Re_*+3.47=18.18$.  Thus, the ratio $U_\mathrm{max}/U_\mathrm{mean}=1.19$ is obtained, which is close to the value of 1.16 obtained in the direct numerical simulation by Kim \textit{et al.}~\cite{Kim}.  The skin-friction coefficient, $C_\mathrm{f}=2u_*^2/U_\mathrm{mean}^2=8.44\times 10^{-3}$ ($8.18\times 10^{-3}$ in the direct numerical simulation by Kim \textit{et al.}).  This value also fits the correlation of Dean~\cite{dean1978reynolds}, who suggested a value of $C_\mathrm{f}=0.073 Re_\mathrm{mean}^{-0.25}$, corresponding to $0.073\times 5542^{-0.25}=8.46\times 10^{-3}$ in the present simulation.

\begin{figure}
	\centering
		\subfigure[]{\includegraphics[width=0.45\textwidth]{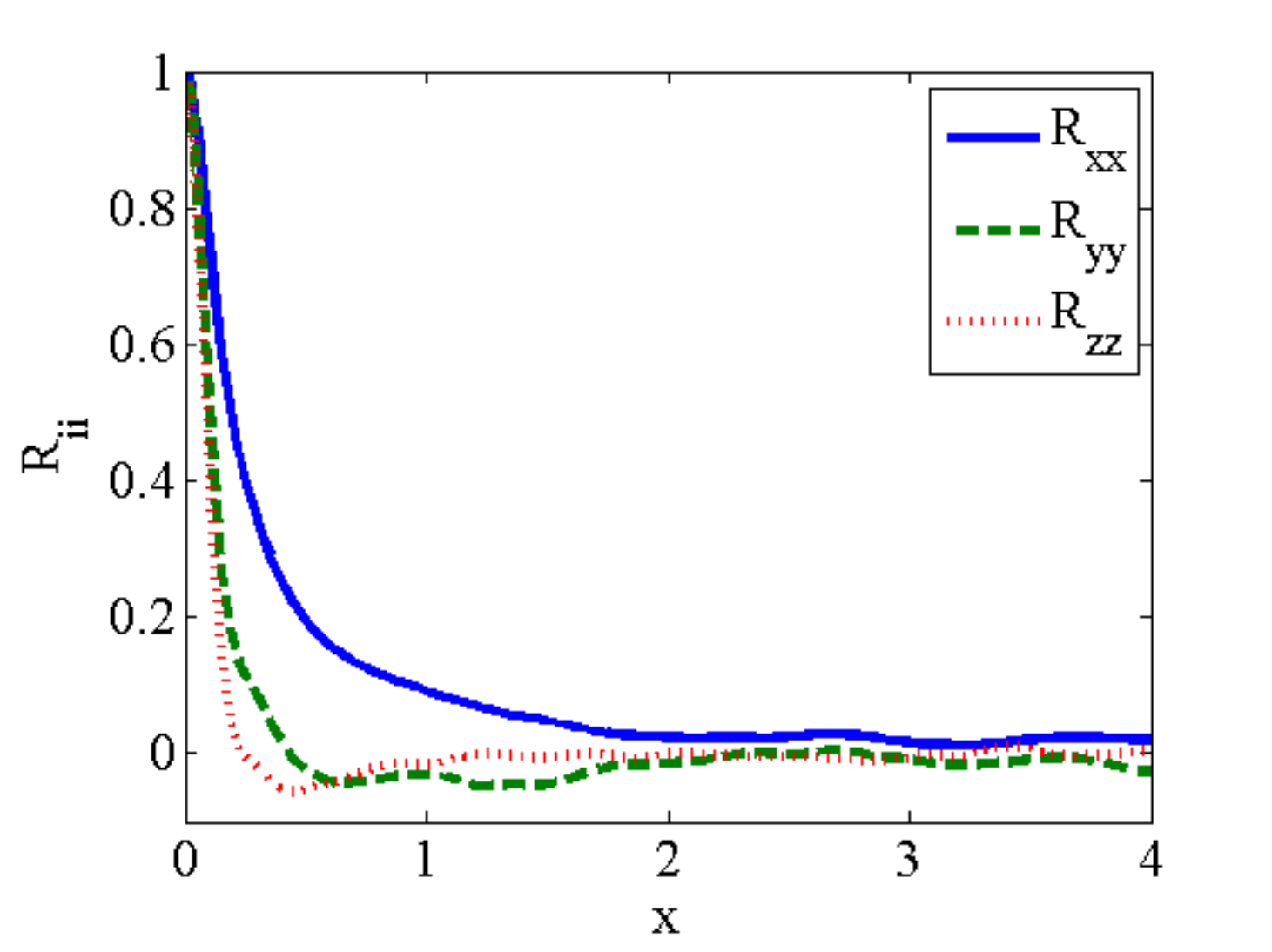}}
		\subfigure[]{\includegraphics[width=0.45\textwidth]{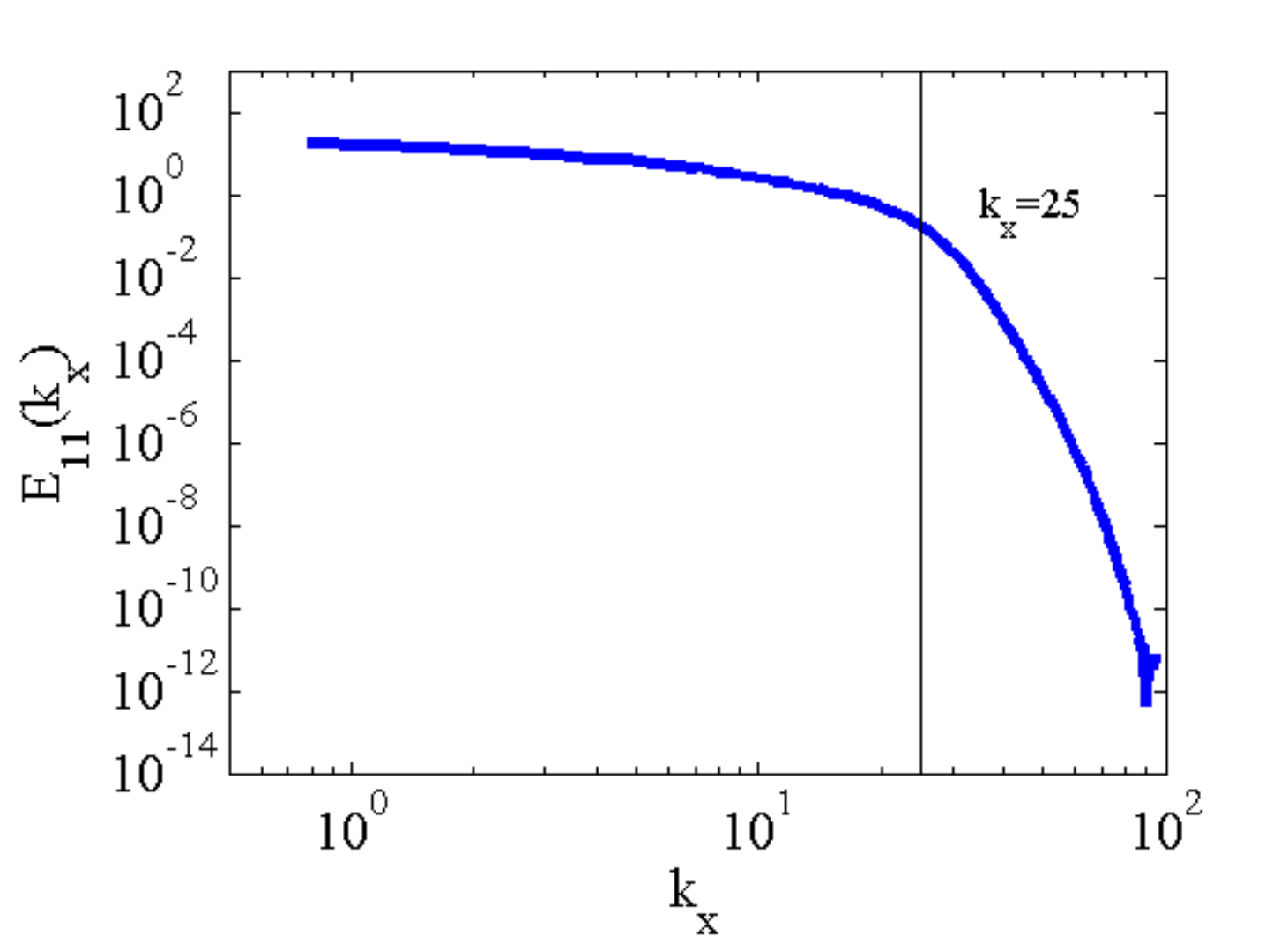}}
		\caption{(a) Streamwise two-point correlation functions and (b) Fourier transform of the streamwise correlation $R_{xx}$.  The onset of the dissipation range is shown in (b).}
	\label{fig:corr_ii_x_shortwindow}
\end{figure}

We examine the following two-point correlation functions
\[
R_{ij}(x,y,z)=\frac{1}{t_2-t_1}\frac{1}{L_xL_y}\int_{t_1}^{t_2}\mathd t\int_0^{L_x}\mathd x'\int_0^{L_y}\mathd y'u_i(x+x',y+y',z)u_j(x',y',z) - \langle u_i\rangle(z)\langle u_j\rangle (z);
\]
in particular we investigate the streamwise correlations $R_{ij}(x):=R_{ij}(x,0,1/2)$, the results of which are shown in Figure~\ref{fig:corr_ii_x_shortwindow}(a).  The two-point correlations vanish at large separations, which confirms that the choice of $L_x=8$ is adequate for the simulation.
To further understand the distribution of the turbulent kinetic energy among the different length scales we introduce the Fourier transform of $R_{ij}(x,y,z)$ in a plane parallel to the walls at location $z$, defined as follows:
\[
\hat{R}_{ij}(k_x,k_y,z)=\iint_{[0,L_x]\times[0,L_y]}\mathd x\mathd y \,\mathe^{-\imag k_x x-\imag k_y y}R_{ij}(x,y,z).
\]
We examine $E_{11}(k_x,z):=|\hat{R}_{ij}(k_x,0,z)|$ at the channel midpoint $z=1/2$ in Figure~\ref{fig:corr_ii_x_shortwindow}(b).  The famous Kolmogorov scaling law $E_{11}(k_x)\sim k_x^{-5/3}$  valid asymptotically at high Reynolds numbers is not in evidence but this is not surprising, as the present simulation is performed at $Re_*=360$, which as at the low end for a turbulent flow.  Indeed, the results are qualitatively very similar to the observed and model results for low-Reynolds number turbulence, in particular the data and model curves for the universal one-dimensional longitudinal velocity spectra given in Reference~\cite{Pope} (Chapter 6 therein).  The spectrum falls off in exponential-type manner for large wavenumbers, corresponding to the dissipation range.  More precisely, the onset of the dissipation range is expected for $k\eta\approx 0.3$, where $\eta$ is the Kolmogorov scale.  The Kolmogorov lengthscale is calculated from
%
%
%
$\eta=(\nu^3/\epsilon)^{1/4}$, where $\epsilon=Re_*^{-1}\langle s_{ij}s_{ij}\rangle$ is the dissipation rate of turbulent kinetic energy and where $s_{ij}$ is the fluctuating component of the rate-of-strain tensor.  For the present simulation, the Kolmogorov lengthscale is calculated to be approximately two wall units, hence $\eta=2/Re_*$ in non-dimensional terms.  In contrast, in Figure~\ref{fig:corr_ii_x_shortwindow}(b) one can see that the onset of the dissipation range occurs at $k\eta\approx 2\times 25/Re_*\approx 0.1$.  Yet again, this is consistent with our knowledge of LES: by design, the Smagorinsky model introduces extra dissipation in the small scales (yet starting at scales larger than the Kolmogorov lengthscale), so the early onset of the dissipation range in Figure~\ref{fig:corr_ii_x_shortwindow}(a) is expected.  Summarizing, the present study confirms that LES is highly effective in capturing the bulk statistics quantitatively very well (e.g. the spatial structure of the mean flow, the Reynolds stress, and the turbulent kinetic energy in the wall-normal direction), yet only captures the fine-scale statistics in a qualitative manner, albeit that the spectra emanating from the LES can be put on a very firm theoretical footing~\cite{Pope}.


\section{Conclusions and didactic considerations}
\label{sec:conc}

Summarizing, the modified and simplified \simples solver introduced in this work performs well in simulating single-phase pressure-driven channel flow.  The code reproduces the Poiseuille base state, and accurately predicts the instability of the same, at sufficiently high Reynolds numbers.  For the instability, excellent agreement between the numerical simulations and Orr--Sommerfeld linear stability theory is obtained.  Following on from this, when introduced to the \simples code, the LES model accurately describes turbulence phenomena in a channel flow, as evidenced by a rigorous comparison between our own simulation results and the standard results from the literature.

This approach would be useful in a learning context for several reasons: TPLS and the \simples derivative are free open-source software and both can be downloaded from a repository -- see References~\cite{s-tpls,tpls_sourceforge}, such that it is readily available to students for immediate use.  Its open-source nature means that students can familiarize themselves with the standard algorithms implemented in the code.  A further advantage is that the code is fully parallelized in a simple but robust way, meaning that the code can be implemented on a wide variety of platforms, from desktops to compute clusters, even up to implementation on supercomputers.  Since the code uses only standard MPI and Fortran, it is highly portable, thereby further enabling easy access by students.  A further final advantage of open-source software is the ability of users to freely modify the source code.  Therefore, the present implementation of TPLS can be used  not only for study and instruction but also as a starting-point for research into temporally-evolving three-dimensional flows.

\subsection*{Acknowledgements}

L\'ON and JF acknowledge the DJEI/DES/SFI/HEA Irish Centre for High-End Computing (ICHEC) for the provision of computational facilities and support.  L\'ON and JF also acknowledge Hendrik Hoffmann's maintenance of the Orr computer cluster in the School of Mathematical Sciences in UCD and support for users of the same.

\bibliographystyle{unsrt}

\begin{thebibliography}{10}

\bibitem{scott2013}
D.~M. Scott, I.~Bethune, L.~\'O N\'araigh, and P.~Valluri.
\newblock Performance enhancement and optimization of the tpls and dim
  two-phase flow solvers.
\newblock Technical report, HECToR dCSE Report, 2013.
\newblock
  \texttt{http://www.hector.ac.uk/cse/distributedcse/reports/tpls-dim/tpls-dim.pdf}.

\bibitem{tpls_sourceforge}
{TPLS: High Resolution Direct Numerical Simulation of Two-Phase Flows}.
\newblock http://sourceforge.net/projects/tpls/.

\bibitem{lennon_1}
L.~\'O N\'araigh, P.~Valluri, D.~M. Scott, I.~Bethune, and P.~D.~M. Spelt.
\newblock Linear instability, nonlinear instability and ligament dynamics in
  three-dimensional laminer two-layer liquid-liquid flows.
\newblock {\em J. Fluid Mech.}, 750:464--506, 2014.

\bibitem{petsc}
Satish Balay, Shrirang Abhyankar, Mark~F. Adams, Jed Brown, Peter Brune, Kris
  Buschelman, Lisandro Dalcin, Victor Eijkhout, William~D. Gropp, Dinesh
  Kaushik, Matthew~G. Knepley, Lois~Curfman McInnes, Karl Rupp, Barry~F. Smith,
  Stefano Zampini, and Hong Zhang.
\newblock {PETS}c {W}eb page.
\newblock http://www.mcs.anl.gov/petsc, 2015.

\bibitem{netcdf}
{NetCDF} web page.
\newblock http://www.unidata.ucar.edu/software/netcdf.

\bibitem{bethune2015}
I.~Bethune, T.~Collis, L.~\'O N\'araigh, D.~Scott, and P.~Valluri.
\newblock {Developing a scalable and flexible high-resolution DNS code for
  two-phase flows}.
\newblock In {\em Proceedings of International Conference on Parallel Computing
  (ParCo) 2015}, September 2015.
\newblock In press.

\bibitem{berselli}
Luigi Berselli, Traian Iliescu, and William~J Layton.
\newblock {\em Mathematics of large eddy simulation of turbulent flows}.
\newblock Springer Science \& Business Media, 2005.

\bibitem{Abbott}
Michael~Barry Abbott and David~R Basco.
\newblock Computational fluid dynamics-an introduction for engineers.
\newblock {\em NASA STI/Recon Technical Report A}, 90:51377, 1989.

\bibitem{CMI}
{Existence and smoothness of the Navier--Stokes equation}.
\newblock \url{http://www.claymath.org/sites/default/files/navierstokes.pdf}.

\bibitem{Pope}
Stephen~B Pope.
\newblock {\em Turbulent flows}.
\newblock Cambridge university press, 2000.

\bibitem{Kim}
John Kim, Parviz Moin, and Robert Moser.
\newblock Turbulence statistics in fully developed channel flow at low reynolds
  number.
\newblock {\em Journal of fluid mechanics}, 177:133--166, 1987.

\bibitem{Deardoff}
James~W Deardorff.
\newblock A numerical study of three-dimensional turbulent channel flow at
  large reynolds numbers.
\newblock {\em Journal of Fluid Mechanics}, 41(02):453--480, 1970.

\bibitem{Davidson}
Peter Davidson et~al.
\newblock {\em Turbulence: an introduction for scientists and engineers}.
\newblock Oxford University Press, USA, 2015.

\bibitem{VanDriest}
Edward~R Van~Driest.
\newblock On turbulent flow near a wall.
\newblock {\em Journal of the Aeronautical Sciences (Institute of the
  Aeronautical Sciences)}, 23(11), 2012.

\bibitem{kang2000boundary}
Myungjoo Kang, Ronald~P Fedkiw, and Xu-Dong Liu.
\newblock A boundary condition capturing method for multiphase incompressible
  flow.
\newblock {\em Journal of Scientific Computing}, 15(3):323--360, 2000.

\bibitem{Harlow}
Francis~H Harlow, J~Eddie Welch, et~al.
\newblock Numerical calculation of time-dependent viscous incompressible flow
  of fluid with free surface.
\newblock {\em Physics of fluids}, 8(12):2182, 1965.

\bibitem{Chorin}
Alexandre~Joel Chorin.
\newblock Numerical solution of the navier-stokes equations.
\newblock {\em Mathematics of computation}, 22(104):745--762, 1968.

\bibitem{Boyd}
John~P Boyd.
\newblock {\em Chebyshev and Fourier spectral methods}.
\newblock Courier Corporation, 2001.

\bibitem{garcia2000}
Alejandro~L Garcia.
\newblock {\em Numerical Methods for Physics}.
\newblock Prentice Hall, second edition, 2000.

\bibitem{shu1988efficient}
Chi-Wang Shu and Stanley Osher.
\newblock Efficient implementation of essentially non-oscillatory
  shock-capturing schemes.
\newblock {\em Journal of Computational Physics}, 77(2):439--471, 1988.

\bibitem{liu1994weighted}
Xu-Dong Liu, Stanley Osher, and Tony Chan.
\newblock Weighted essentially non-oscillatory schemes.
\newblock {\em Journal of computational physics}, 115(1):200--212, 1994.

\bibitem{durran1991third}
Dale~R Durran.
\newblock The third-order adams-bashforth method: An attractive alternative to
  leapfrog time differencing.
\newblock {\em Monthly weather review}, 119(3):702--720, 1991.

\bibitem{Yang}
Xiyang~I.A. Yang and Rajat Mittal.
\newblock Acceleration of the jacobi iterative method by factors exceeding 100
  using scheduled relaxation.
\newblock {\em Journal of Computational Physics}, 274:695 -- 708, 2014.

\bibitem{orr}
William~M'F. Orr.
\newblock The stability or instability of the steady motions of a perfect
  liquid and of a viscous liquid. part ii: A viscous liquid.
\newblock {\em Proceedings of the Royal Irish Academy. Section A: Mathematical
  and Physical Sciences}, 27:69--138, 1907.

\bibitem{valluri2010linear}
P~Valluri, L~{\'O} N{\'a}raigh, H~Ding, and PDM Spelt.
\newblock Linear and nonlinear spatio-temporal instability in laminar two-layer
  flows.
\newblock {\em Journal of Fluid Mechanics}, 656:458--480, 2010.

\bibitem{schmid2012stability}
Peter~J Schmid and Dan~S Henningson.
\newblock {\em Stability and transition in shear flows}, volume 142.
\newblock Springer Science \& Business Media, 2012.

\bibitem{Orszag}
Steven~A Orszag.
\newblock Accurate solution of the orr--sommerfeld stability equation.
\newblock {\em Journal of Fluid Mechanics}, 50(04):689--703, 1971.

\bibitem{drazin2004hydrodynamic}
Philip~G Drazin and William~Hill Reid.
\newblock {\em Hydrodynamic stability}.
\newblock Cambridge university press, 2004.

\bibitem{batten2004interfacing}
Paul Batten, Uriel Goldberg, and Sukumar Chakravarthy.
\newblock Interfacing statistical turbulence closures with large-eddy
  simulation.
\newblock {\em AIAA journal}, 42(3):485--492, 2004.

\bibitem{benocci2006large}
C~Benocci, JPAJ van Beeck, and Ugo Piomelli.
\newblock {\em Large eddy simulation and related techniques: theory and
  applications: March 13-16, 2006}.
\newblock Von Karman Institute for Fluid Dynamics, 2006.

\bibitem{lesieur2005large}
Marcel Lesieur, Olivier M{\'e}tais, and Pierre Comte.
\newblock {\em Large-eddy simulations of turbulence}.
\newblock Cambridge University Press, 2005.

\bibitem{dean1978reynolds}
RB~Dean.
\newblock Reynolds number dependence of skin friction and other bulk flow
  variables in two-dimensional rectangular duct flow.
\newblock {\em Journal of Fluids Engineering}, 100(2):215--223, 1978.

\bibitem{s-tpls}
{S-TPLS source code}.
\newblock https://sourceforge.net/p/tpls/code/HEAD/tree/trunk/s-tpls/.

\end{thebibliography}

\end{document}